\documentclass[12pt]{article}

\usepackage{graphics,graphicx,fullpage,multirow}
\usepackage{amsmath,amssymb,verbatim,natbib}
\usepackage[dvipsnames,usenames]{color}
\usepackage{subfig}
\usepackage{color}

\newtheorem{defin}{\bf Definition}
\newenvironment{definition}{\begin{defin}\rm}{\end{defin}}

\def\ga{\mbox{Ga}}

\def\no{\mbox{N}}

\def\dir{\mbox{Dir}}

\def\E{\mbox{E}}
\def\V{\mbox{Var}}

\def\P{\mbox{P}}

\def\d{\mbox{d}}
\def\data{\mbox{data}}

\def\br{{\bf r}}

\def\bx{{\bf x}}
\def\by{{\bf y}}
\def\bz{{\bf z}}
\def\bH{{\bf H}}
\def\bI{{\bf I}}
\def\bN{{\bf N}}
\def\bT{{\bf T}}
\def\bU{{\bf U}}
\def\bX{{\bf X}}
\def\bY{{\bf Y}}
\def\bZ{{\bf Z}}
\def\bzero{{\bf 0}}

\newcommand{\balpha}{\boldsymbol{\alpha}}

\newcommand{\bfeta}{\boldsymbol{\eta}}
\newcommand{\bepsilon}{\boldsymbol{\epsilon}}

\newcommand{\bmu}{\boldsymbol{\mu}}

\newcommand{\btheta}{\boldsymbol{\theta}}
\newcommand{\bTheta}{\boldsymbol{\Theta}}
\newcommand{\bj}{\boldsymbol{j}}

\newcommand{\NB}{\mathbb{N}}
\newcommand{\RB}{\mathbb{R}}
\newcommand{\SB}{\mathbb{S}}

\newcommand{\AC}{\mathcal{A}}
\newcommand{\BC}{\mathcal{B}}

\newcommand{\PPT}{\mbox{PPT}}

\begin{document}

\baselineskip=24pt

\title{{\bf Multivariate and regression models for directional data based on projected P\'olya trees}}
\author{
  {\sc  Luis E. Nieto-Barajas} \\
  {\sl {\small Department of Statistics, ITAM, Mexico}}
  }
\date{}

\maketitle

\begin{abstract}
Projected distributions have proved to be useful in the study of circular and directional data. Although any multivariate distribution can be used to produce a projected model, these distributions are typically parametric. In this article we consider a multivariate P\'olya tree on $\RB^k$ and project it to the unit hypersphere $\SB^k$ to define a new Bayesian nonparametric model for directional data. We study the properties of the proposed model and in particular, concentrate on the implied conditional distributions of some directions given the others to define a directional-directional regression model. We also define a multivariate linear regression model with P\'olya tree errors and project it to define a linear-directional regression model. We obtain the posterior characterisation of all models via their full conditional distributions. Metropolis-Hastings steps are required, where random walk proposal distributions are optimised with a novel adaptation scheme. We show the performance of our models with simulated and real datasets. 
\end{abstract}


\section{Introduction}
\label{sec:intro}

Directional data arise from the observation of unit vectors in a $k$-dimensional space, $\SB^{k}$, which  is also called $k$-dimensional unit hypersphere. These data can also be represented as $k-1$ angles in polar coordinates, also known as direction cosines. Particular cases arise when $k=2$ producing circular data, and when $k=3$ producing spherical data. If data are only observed in a portion of a hypersphere, they are known as axial data. In general, directional data occur in different disciplines like biology, geophysics, meteorology, ecology and environmental sciences. Specific applications include the study of wind directions, orientation data in biology, direction of birds migration, directions of fissures propagation in concrete and other materials, orientation of geological deposits, and the analysis of mammalian activity patterns in ecological reserves, among others. For a survey on the area, the reader is referred to \cite{mardia:72}, \cite{fisher&al:87} and \cite{mardia&jupp:99}. 

The study of directional data started in the early years of the XX century. For instance, the von Mises distribution \citep{vonmises:18}, wrapped distributions \citep{levy:39} and projected, offset or displaced distributions \citep{klotz:64}. In the XXI century, thanks to the advances of the statistical science, new more elaborate models have been proposed, specially in the context of Bayesian nonparametrics. For instance, \cite{nunez&al:15} proposed a Dirichlet process mixture of wrapped and projected normals, and \cite{nieto&nunez:21} introduced a projected P\'olya tree, both for the analysis of circular data. 

Following \cite{jamma&gupta:01}, regression models for circular response variables can be classified according to whether the explanatory variables are linear (linear-circular) or circular (circular-circular). For the linear-circular case \cite{gould:69} considered a von Mises distribution for the response with mean parameter expressed in terms of a linear predictor of explanatory variables. Later \cite{fisher&lee:92} generalised the previous model by including a link function of the explanatory variables and they further proposed a concentration parameter regression. \cite{presnell&al:98} considered a multivariate normal regression model with linear responses and linear predictors and projected the responses to the unit hypersphere. For the circular-circular case, \cite{jamma&sarma:93} defined the conditional first trigonometric moment of the response in terms of functions of the explanatory variable and \cite{downs&mardia:02} considered a von Mises distribution for the response and proposed a particular link function of a circular explanatory variable. 

The unit hypersphere $\SB^k$ can also be seen as a non-euclidean manifold. In this more general setting, \cite{gutierrez&al:19} proposed a Bayesian approach for shape analysis using the projected normal distribution. \cite{bhattacharya&dunson:10} and \cite{lin&al:17}, on the other hand, proposed Dirichlet process mixture models with suitable kernels on planar shape spaces and on the stiefel manifold, respectively. 

The idea of this article is to introduce a multivariate model for directional data of any dimension $k$ based on a projected P\'olya tree. To the best of our knowledge this would be the first Bayesian nonparametric model specifically designed for directional ($k\geq 3$) data. By considering the implied conditional distributions we also define directional-directional regression models. If before projection we consider a multivariate median regression model with P\'olya tree errors and with linear explanatory variables, the projected model for the responses is a linear-directional regression model. Posterior inference is carried out via the full conditional distributions. To sample from them, several Metropolis-Hastings (MH) steps are required, where random walk proposal distributions are optimised with a novel adaptation scheme. 

To place our contribution in proper context, the projected P\'olya tree of \cite{nieto&nunez:21} considers a bivariate P\'olya tree and projects it to the unit circle for the analysis of circular data in $[0,2\pi]$. Extension of the bivariate P\'olya tree to a multivariate P\'olya tree could be seen, in principle, a straightforward task. However, there are many possible partitions that could be used to define the tree. Here we propose an orthogonal partition of $\RB^k$. Second, projected models for directional data in $\SB^k$ for $k\geq 3$ are challenging since the support of the $k-1$ vector of angles is not simply $[0,2\pi]^{k-1}$ but $[0,\pi]^{k-2}\times[0,2\pi]$, that is, $k-2$ axial dimensions plus a circular dimension. Additionally, Bayesian nonparametric regression models for directional data have not been studied before. Showing these and other differences is the purpose of this work.

The rest of the paper is organized as follows. In Section \ref{sec:mpt} we set the notation and present basic ideas about multivariate P\'olya trees. In Section \ref{sec:model} we introduce the directional model as a projected P\'olya tree prior and define the two types of regression models. In Section \ref{sec:post} we describe how to perform posterior inference via a data augmentation technique,  consider hyper priors for some of the parameters and propose a novel adaptation scheme. We illustrate the performance of our proposal in Section \ref{sec:numerical}, for the spherical case ($k=3$), via simulations and the analysis of real datasets with and without covariates. We conclude with some remarks in Section \ref{sec:concl}.

Before proceeding we introduce notation: $\ga(\alpha,\beta)$ denotes a gamma density with mean $\alpha/\beta$; $\no(\mu,\tau)$ denotes a normal density with mean $\mu$ and precision $\tau$; $\no_k(\bmu,\bT)$ denotes a $k$-variate normal density with mean vector $\bmu$ and precision matrix $\bT$; and $\dir(\balpha)$ denotes a dirichlet density with parameter vector $\balpha$.

\section{Multivariate P\'olya tree}
\label{sec:mpt}

In this section we recall the definition of a multivariate P\'olya tree and set notation. Let $(\mathbb{R}^k,\BC^k)$ be a measurable space for $k\in\NB$. There are several ways of defining and denoting the nested partition $\Pi$ \citep{hanson:06,jara&al:09,filippi&holmes:17}. For simplicity, we define the partition as the cross product of univariate partitions and use the notation of \cite{nieto&mueller:12}. In other words, $\Pi=\{B_{m,\bj_m}\}$ such that $B_{m,\bj_m}=B_{m,j_{m,1}}\times\cdots\times B_{m,j_{m,k}}$, for $j_{m,l}=1,\ldots,2^{m}$, $l=1,\ldots,k$ and $m=1,2,\ldots$. The index $m$ denotes the level of the tree and the vector $\bj_m=(j_{m,1},\ldots,j_{m,k})$ locates the partitioning subset within the level. In general, the set $B_{m,\bj_m}$ splits into $2^k$ disjoint subsets $B_{m+1,\bj_{m+1}}$ such that $\bj_{m+1}\in\{\times_{l=1}^k(2j_{m,l}-1,2j_{m,l})\}$. At each level $m$ we will have a partition of size $2^{km}$. We associate random branching probabilities $Y_{m,\bj_m}$ with every set $B_{m,\bj_m}$ such that, for example, $Y_{m+1,2j_1-1,\ldots,2j_k-1}=F(B_{m+1,2j_1-1,\ldots,2j_k-1}\mid B_{m,\bj_m})$, where $F$ denotes a cumulative distribution function or a probability measure, indistinctively.

\begin{definition}
\label{def:mPT}
Let $\AC=\{\alpha_{m,\bj_m}\}$, $\bj_m=(j_{m,1},\ldots,j_{m,k})$, $j_{m,l}=1,\ldots,2^m$, $l=1,\ldots,k$, $m=1,2,\ldots$ be a set of nonnegative real numbers. A random probability measure $F$ on $(\RB^k,\BC^k)$ is said to have a multivariate P\'olya tree prior with parameters $(\Pi,\AC)$ if there exists random vectors $\bY_{m,\bj_m}=\{Y_{m+1,\bj_{m+1}},\,\bj_{m+1}\in\{\times_{l=1}^k(2j_{m,l}-1,2j_{m,l})\}\}$ of dimension $2^k$ such that the following hold: 
\begin{enumerate}
\item All random vectors $\bY_{m,\bj_m}$, for $\bj_m=(j_{m,1},\ldots,j_{m,k})$, $j_{m,l}=1,\ldots,2^m$, $l=1,\ldots,k$, $m=0,1,\ldots$ are independent
\item For every $m=0,1,\ldots$ and every $\bj_m$, $\bY_{m,\bj_m}\sim\dir(\balpha_{m,\bj_m})$, where \\ $\balpha_{m,\bj_m}=\{\alpha_{m+1,\bj_{m+1}},\,\bj_{m+1}\in\{\times_{l=1}^k(2j_{m,l}-1,2j_{m,l})\}\}$ such that $\dim(\balpha_{m,\bj_m})=2^k$
\item For every $m=1,2,\ldots$ and every $\bj_m$, $$F(B_{m,\bj_m})=\prod_{h=1}^m Y_{m-h+1,j_{m-h+1,1},\ldots,j_{m-h+1,k}},$$ where $j_{h-1,l}=\left\lceil\frac{j_{h,l}}{2}\right\rceil$ for $l=1,\ldots,k$  are recursive decreasing formulae that locate the set $B_{m,\bj_m}$ with its ancestors upwards in the tree. 
\end{enumerate}
\end{definition}
In Definition \ref{def:mPT}, $\bY_{0,1,\ldots,1}=\{Y_{1,\bj_1},\bj_1\in\{\times_{l=1}^k(1,2)\}\}$ and $\balpha_{0,1,\ldots,1}=\{Y_{1,\bj_1},\bj_1\in\{\times_{l=1}^k(1,2)\}\}$ are the vectors associated to the partition elements at level $m=1$.

We can center the multivariate P\'olya tree around a parametric probability measure $F_0$. For simplicity, let us assume that $F_0(x_1,\ldots,x_k)=\prod_{l=1}^kF_{0,l}(x_l)$. Non-independence $F_0$ could also be considered but a suitable transformation of the partition sets $B_{m,\bj_m}$ would be required \citep[e.g.][]{jara&al:09}. Therefore, we proceed by matching the marginal partitions $\left\{B_{m,j_{m,l}}\right\}$ with the dyadic quantiles of the marginals $F_{0,l}$, i.e.,
\begin{equation}
\label{eq:bmj2}
B_{m,j_{m,l}}=\left(F_{0,l}^{-1}\left(\frac{j_{m,l}-1}{2^m}\right),F_{0,l}^{-1}\left(\frac{j_{m,l}}{2^m}\right)\right],
\end{equation}
for $j_{m,l}=1,\ldots,2^m$ and $l=1,\ldots,k$. We further define $\balpha_{m,\bj_m}=(\alpha\varphi(m+1),\ldots,\alpha\varphi(m+1))$ where $\alpha>0$ can be interpreted as the precision parameter \citep{walker&mallick:97} and the function $\varphi$ controls the speed at which the variance of the branching probabilities moves down in the tree. As suggested by \cite{watson&al:17} we take $\varphi(m)=m^\delta$ with $\delta>1$ to define an absolutely continuous multivariate P\'olya tree \citep{kraft:64}. It is not difficult to prove that a multivariate P\'olya tree, defined in this way, satisfies $\E\{F(B_{m,\bj_m})\}=F_0(B_{m,\bj_m})=(1/2)^{km}$. 

In practice we need to stop partitioning the space at a finite level $M$ to define a finite tree process. At the deepest level $M$, we can spread the probability within each set $B_{M,\bj_M}$ according to $f_0$, the density associated to $F_0$. In this case the random probability measure defined will have a multivariate density of the form
\begin{equation}
\label{eq:dPT}
f(\bx)=\left\{\prod_{m=1}^M Y_{m,\bj_m^{(\bx)}}\right\}2^{kM} f_0(\bx),
\end{equation}
where $\bx'=(x_1,\ldots,x_k)\in\RB^k$, and with $\bj_m^{(\bx)}=(j_{m,1}^{(x_1)},\ldots,j_{m,k}^{(x_k)})$ identifying the set at level $m$ that contains $\bx$. This maintains the condition $\E(f)=f_0$. By taking $M\to\infty$ we recover the (infinite) multivariate P\'olya tree of Definition \ref{def:mPT}.

P\'olya tree densities are discontinuous at the boundaries of the partitions \citep[e.g.][]{lavine:92}. To overcome this feature, an extra mixture with respect to the parameters of the centering measure is imposed, that is, $f_0$ is replaced by $f_0(\cdot\mid\bfeta)$ and a prior $f(\bfeta)$ is placed to induce smoothness.

\section{Projected tree models}
\label{sec:model}

\subsection{Directional and circular models}

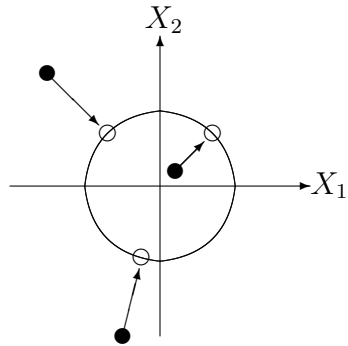
\begin{figure}
\setlength{\unitlength}{1cm}
\begin{center}
\begin{picture}(4,4)
\put(0,2){\vector(1,0){4}}
\put(2,0){\vector(0,1){4}}
\qbezier(1,2)(1.1,2.9)(2,3)
\qbezier(2,3)(2.9,2.9)(3,2)
\qbezier(1,2)(1.1,1.1)(2,1)
\qbezier(2,1)(2.9,1.1)(3,2)
\put(2.2,2.2){\circle*{0.2}}
\put(2.2,2.2){\vector(1,1){0.39}}
\put(2.7,2.7){\circle{0.2}}
\put(1.5,0){\circle*{0.2}}
\put(1.5,0){\vector(1,4){0.225}}
\put(1.75,1.05){\circle{0.2}}
\put(0.5,3.5){\circle*{0.2}}
\put(0.5,3.5){\vector(1,-1){0.69}}
\put(1.3,2.7){\circle{0.2}}
\put(4,1.9){$X_1$}
\put(1.8,4.1){$X_2$}
\end{picture}
\end{center}
\caption{Graphical representation of full dots in $\RB^2$ projected to empty dots in $\SB^2$.}
\label{fig:proj} 
\end{figure}

We are now in a position to construct the projected P\'olya tree. Let us assume a multivariate random vector $ \bX'=(X_1,\ldots,X_k)$ such that $\bX\mid f \sim f$ and $f$ is given in \eqref{eq:dPT}. We project the random vector $\bX$ to the unit circle by defining $\bU=\bX/R$, where $R=||\bX||$ is the resultant length of the vector. Figure \ref{fig:proj} depicts a graphical representation of the projection for $k=2$. Alternatively, we can work with the multivariate polar coordinates transformation $\bX\leftrightarrow (\bTheta,R)$, where $\bTheta=(\Theta_1,\ldots,\Theta_{k-1})$ are the $k-1$ angles such that $\btheta\in\bH$ with $\bH=\{\theta_l\in[0,\pi],l=1,\ldots,k-2,\theta_{k-1}\in[0,2\pi],\}$ and $R\in\RB^+$. From polar to linear variables the transformation is 
\begin{align}
\nonumber
X_1=R\cos\Theta_1,\; X_2=R\sin\Theta_1\cos\Theta_2,\; \ldots,\\ 
\label{eq:xth}
X_{k-1}=R(\prod_{l=1}^{k-2}\sin\Theta_l)\cos\Theta_{k-1},\; X_k=R\prod_{l=1}^{k-1}\sin\Theta_l.
\end{align}
Thus, after some algebra, we obtain the Jacobian $J=(-r)^{k-1}\prod_{l=1}^{k-2}(\sin\theta_l)^{k-l-1}$. Then, the induced marginal density for the $k-1$ angles $\bTheta$ has the form
\begin{equation}
\label{eq:PPT}
f(\btheta)=\int_0^\infty\left\{\prod_{m=1}^M Y_{m,\bj_m^{(\bx(\btheta,r))}}\right\}2^{kM} f_0(\bx(\btheta,r))\,|J|\,\d r,
\end{equation}
where we have made explicitly that $\bx$ has to be written in terms of the multivariate polar coordinates $(\btheta,r)$. We will refer to the density $f(\btheta)$, given in \eqref{eq:PPT}, as the $(k-1)$-directional model based on a projected P\'olya tree, and will be denoted by $\PPT_{k-1}(\alpha,\varphi,f_0,M)$. 

As noted by \cite{nieto&nunez:21}, projected P\'olya tree densities, as in \eqref{eq:PPT}, are not discontinuous at the boundaries of the partitions. This is a consequence of the marginalisation when passing from the joint density $f(\btheta,r)$ to the marginal $f(\btheta)$, which can also be seen as a mixture of the form $f(\btheta)=\int f(\btheta\mid r)f(r)\d r$. A specific angle $\btheta_0$ might come from several points in $\RB^k$ defined by different resultants in polar coordinates say $(\btheta_0,r_h)$, $h=1,2,\ldots$. Each of these points might belong to different partition sets, which are added (integrated) in the marginalisation. 

In particular, we can center our projected P\'olya tree on the projected normal distribution, considered by \cite{wang&gelfand:13}, by taking $f_0(\bx)=\no_k(\bx\mid\bmu,\bI)$, that is, a multivariate normal density with mean vector $\bmu'=(\mu_1,\ldots,\mu_k)$ and precision the $k\times k$ identity matrix $\bI$. In this case, the projected P\'olya tree becomes 
\begin{align}
\nonumber
f(\btheta)=& \,2^{kM} (2\pi)^{-k/2}e^{-\frac{1}{2}\bmu'\bmu}\,\left\{\prod_{l=1}^{k-2}(\sin\theta_l)^{k-l-1}\right\}I_{\bH}(\btheta) \\
\label{eq:PPTn}
&\times\int_0^\infty\left\{\prod_{m=1}^M Y_{m,\bj_m^{\bx(\btheta,r)}}\right\}r^{k-1}\exp\left[-\frac{1}{2}\left\{r^2-2r\bmu'\bx(\btheta,1)\right\}\right]\d r.
\end{align}

Marginal densities for each $\Theta_l$, $l=1,\ldots,k-1$ can be obtained by standard marginalisation procedures. If we define $\bTheta_{-l}=\{\Theta_h:h\neq l\in\{1,\ldots,k-1\}\}$ the vector of all angles except $\Theta_l$, and $\bH_{-l}$ the corresponding support, then marginal densities have the expression 
\begin{equation}
\label{eq:marginal}
f(\theta_l)=\int_{\bH_{-l}}f(\btheta)\d\btheta_{-l}.
\end{equation}
Expression \eqref{eq:marginal} becomes an axial model for $l=1,\ldots,k-2$ and becomes a circular model \citep{nieto&nunez:21} if $l=k-1$. In either case we will denote this model as $\PPT_{1}(\alpha,\varphi,f_0,M)$.

Densities for circular variables on $[0,2\pi]$ are periodic, that is $f(\theta_l+2\pi)=f(\theta_l)$, therefore moments are not defined in the usual way \citep[e.g.][]{jamma&umbach:10}. Instead, the $q^{th}$ trigonometric moment of a random variable $\Theta_l$, $l=1,\ldots,k-1$, is a complex number of the form $\varphi_{l,q}=\E(e^{iq\Theta_l})=a_{l,q}+ib_{l,q}$, for any integer $q$, where $a_{l,q}=\E\{\cos(q\Theta_l)\}$ and $b_{l,q}=\E\{\sin(q\Theta_l)\}$. The mean direction of $\Theta_l$, denoted by $\nu_{l}$, and a concentration measure around the mean, denoted by $\varrho_{l}$, are defined in terms of the moment of order $q=1$ as 
\begin{equation}
\label{eq:moments}
\nu_{l}=\arctan(b_{l,1}/a_{l,1})\quad\mbox{and}\quad\varrho_{l}=\sqrt{a_{l,1}^2+b_{l,1}^2}, 
\end{equation}
where $\varrho_{l}\in[0,1]$ for $l=1,\ldots,k-1$. A value of $\varrho_{l}$ close to one means that $\Theta_l$ is highly concentrated around its mean $\nu_{l}$, whereas a value of $\varrho_{l}$ close to zero means that $\Theta_l$ is highly disperse. 

Association between two random directional variables has been described via generalisations of the Pearson correlation measure. One proposal is that of \cite{jamma&sarma:88} given by 
\begin{equation}
\label{eq:cor}
\rho(\Theta_l,\Theta_h)=\frac{\E\{\sin(\Theta_l-\nu_{l})\sin(\Theta_h-\nu_{h})\}}{\sqrt{\V\{\sin(\Theta_l-\nu_{l})\}\V\{\sin(\Theta_h-\nu_{h})\}}},
\end{equation}
which maintains the properties $|\rho(\Theta_l,\Theta_h)|\leq 1$ and $\rho(\Theta_l,\Theta_h)=0$ if $\Theta_l$ and $\Theta_h$ are independent.

\subsection{Regression models}
\label{sec:reg}

The simple linear regression model for any two (non circular) random variables, say $Y$ and $Z$, arises from a joint normal model for the bivariate vector $f(y,z)$ and defining the conditional distribution of the random variable of interest, say $Y$, given $Z$ of the form $f(y\mid z)$. We extend this idea to directional variables to define a regression model for the variables of interest. Without loss of generality, let $\bTheta_I=\{\Theta_1,\ldots,\Theta_l\}$ be a subset of the $(k-1)$-dimensional vector $\bTheta$, and $\bTheta_{II}=\{\Theta_{l+1},\ldots,\Theta_{k-1}\}$ the complement. If $\bTheta_I$ is the set of responses (variables of interest) and $\bTheta_{II}$ the set of explanatory variables, then the regression model $f(\btheta_I\mid\btheta_{II})$ is given by 
\begin{equation}
\label{eq:reg}
f(\btheta_I\mid\btheta_{II})=\frac{f(\btheta)}{\int_{\bH_I}f(\btheta)\d\btheta_I}
\end{equation}
where $\bH_I$ is the cross product of the corresponding marginal supports $H_l=[0,\pi]$ if $l=1,\ldots,k-2$ or $H_l=[0,2\pi]$ if $l=k-1$; and $f(\btheta)$ is either \eqref{eq:PPT} for the general $\PPT$, or \eqref{eq:PPTn} for the $\PPT$ centred around a projected normal. Expression \eqref{eq:reg} could be seen as a directional-directional regression model emphasising the fact that the response as well as the explanatory variables are directional.  

Linear-directional regression models can also be defined with projected trees. In particular, we consider a median regression model for the unprojected variables $\bX$ given a $p$-set of (linear) explanatory variables $\bZ'=(Z_1,\ldots,Z_p)$, where one of them could be an intercept, of the form 
\begin{equation}
\label{eq:medreg}
\bX=\Gamma\,\bZ+\bepsilon,
\end{equation}
with $\Gamma$ a $(k\times p)$-matrix of coefficients and $\bepsilon'=(\epsilon_1,\ldots,\epsilon_k)$ random errors such that $\bepsilon\mid f\sim f$ where $f$ is a finite multivariate P\'olya tree process, as in \eqref{eq:dPT}, centred around $f_0(\bx)=N_k(\bx\mid\bzero,\bI)$. To make the location $\Gamma\bZ$ identifiable in model \eqref{eq:medreg}, we need to constrain the distribution of the errors to have median zero. This is achieved by imposing an additional constraints in the first level of the tree \citep{hanson&johnson:02}. That is, in \eqref{eq:bmj2}, for $m=1$ and $l=1,\ldots,k$, we require $B_{1,1}=(-\infty,0]$ and $B_{1,2}=(0,\infty)$, together with fixed branching probabilities $\bY_{0,1,\ldots,1}=(1/2^{k},\ldots,1/2^{k})$. These constraints ensure that the marginal medians of $\bX$ are $\Gamma\,\bZ$. If we finally consider the polar coordinates transformation $\bepsilon\leftrightarrow(\Theta,R)$ and marginalise with respect to $R$, the regression model $f(\btheta\mid\bz)$ is of the form \eqref{eq:PPTn} with $\bx(\btheta,r)$ replaced by $\bepsilon(\btheta,r)=\bx(\btheta,r)-\Gamma\,\bz$ and with $\bmu=\bzero$. 

\begin{figure}
\centerline{\includegraphics[scale=0.295]{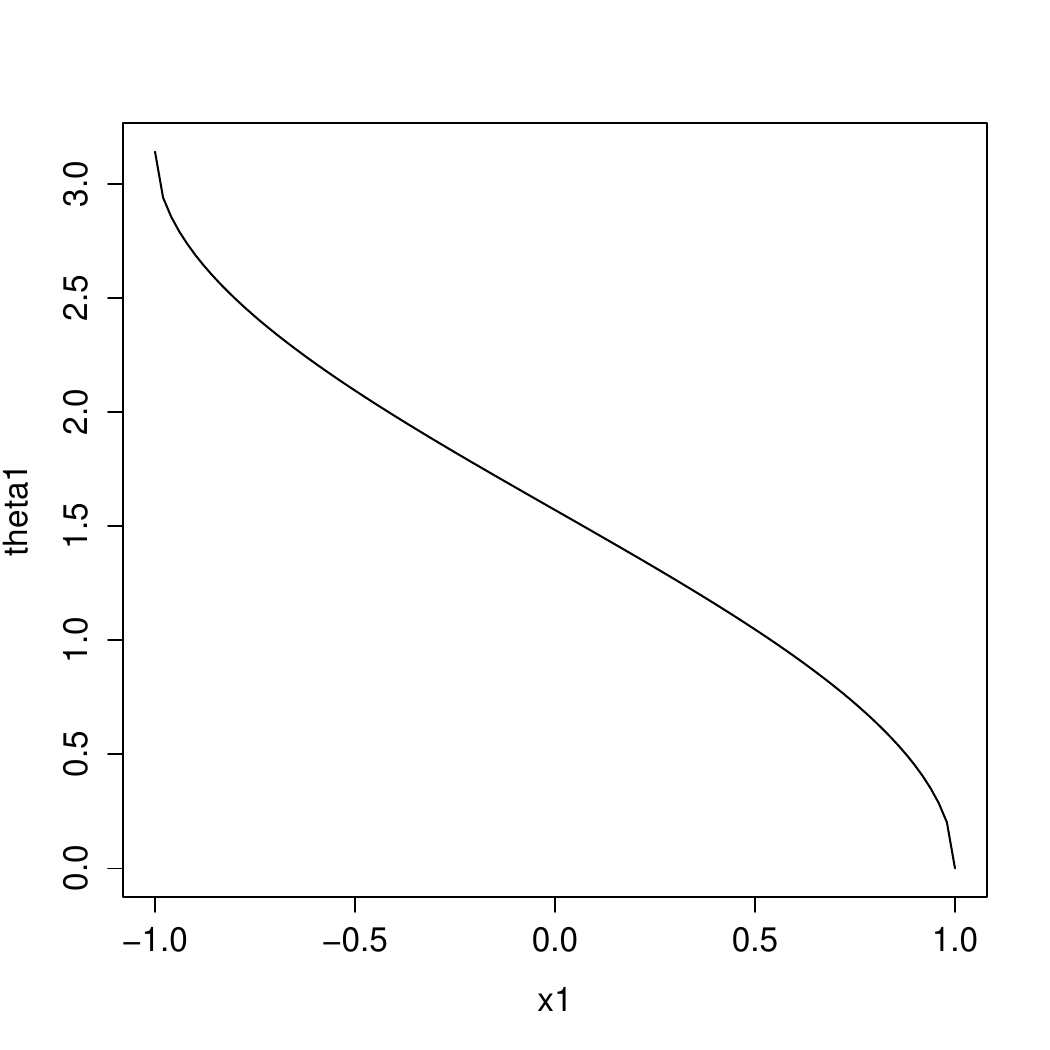}
\includegraphics[scale=0.295]{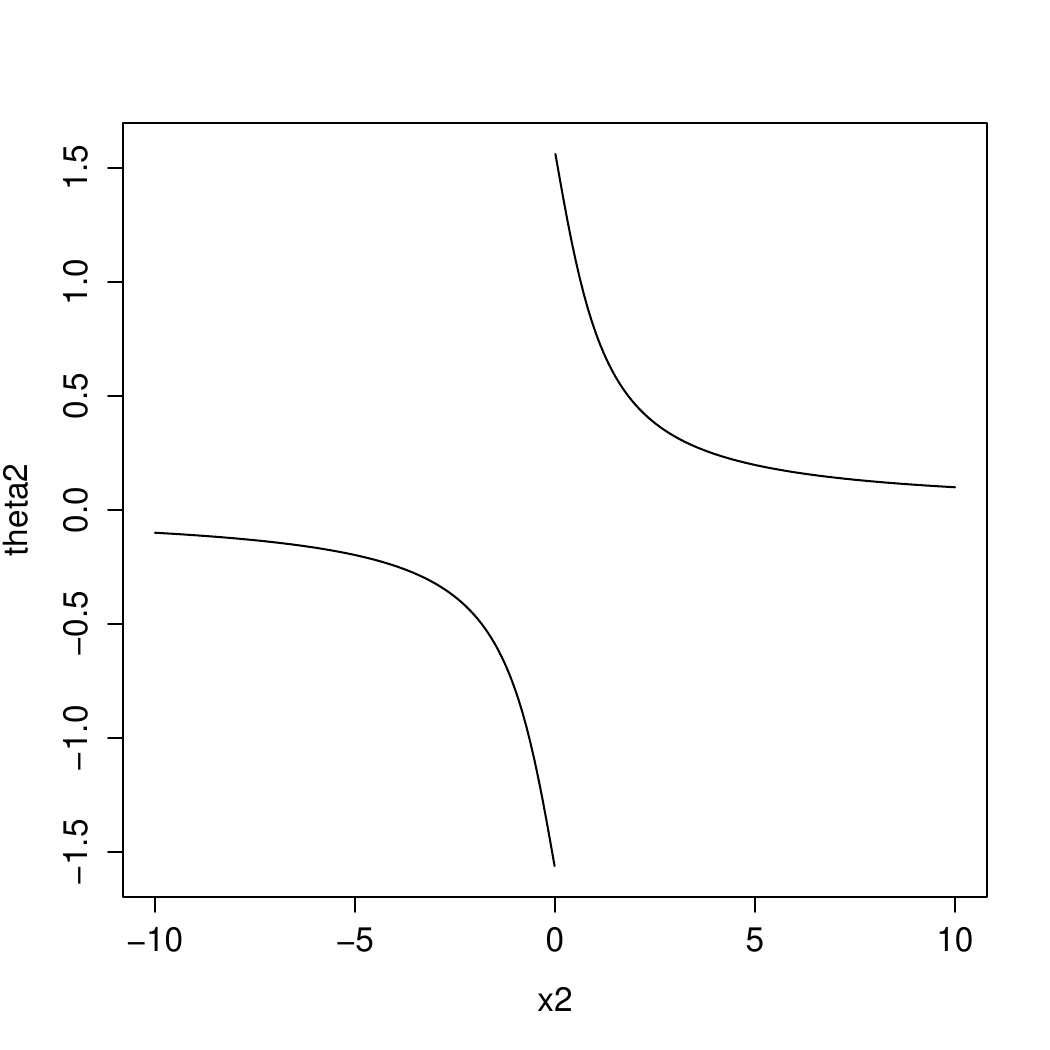}
\includegraphics[scale=0.295]{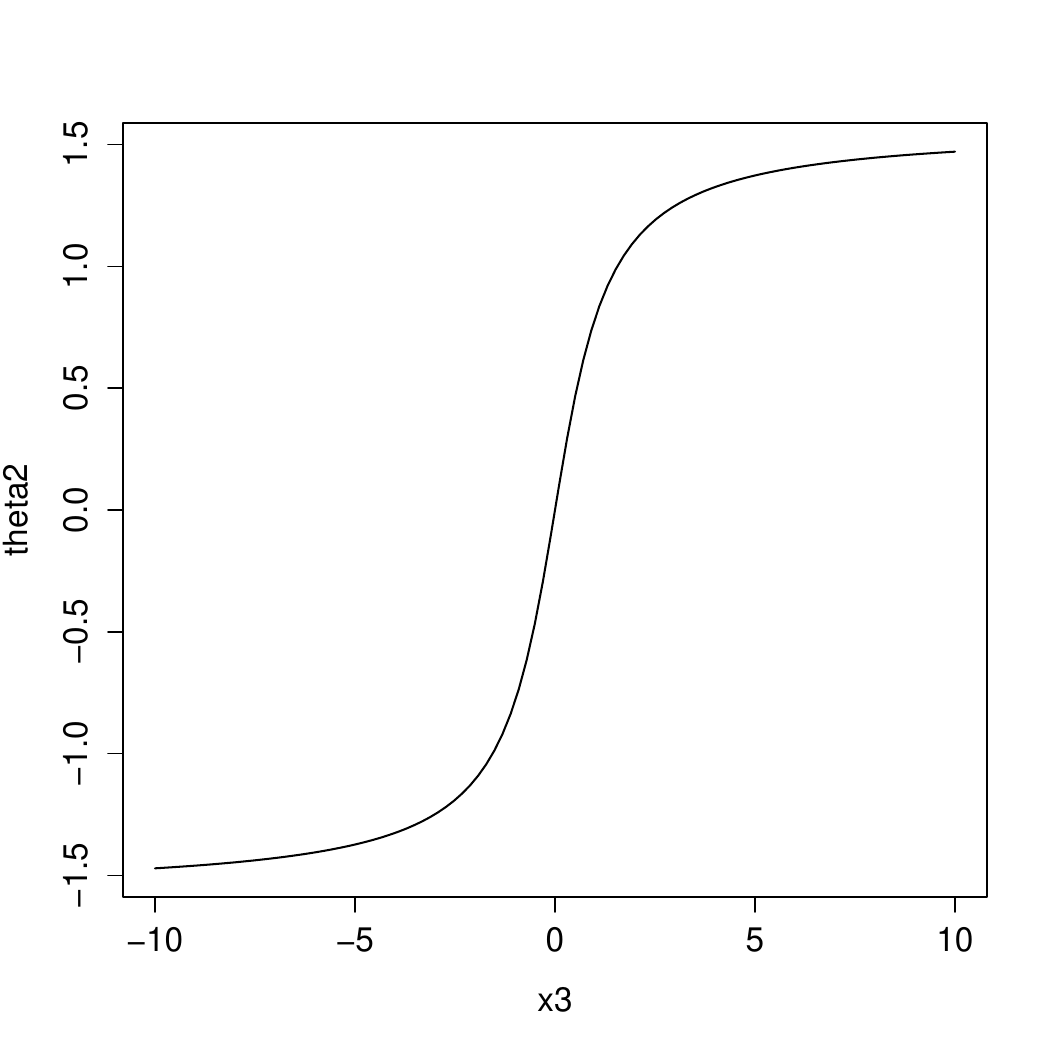}}
\caption{{\small Cosines $\theta_l$ as function of latent $X_h$ for spherical setting ($k=3$), $l=1,2$, $h=1,2,3$.}}
\label{fig:coef}
\end{figure}

Interpretation of $\Gamma=(\gamma_{l,h})$ coefficients on the response $\bTheta$ in the linear-directional regression model might not be direct as it is for the unprojected variables $\bX$. To help with the interpretation, we can rely on graphs of $\Theta_l$ as a function of $X_h$ for $l=1,\ldots,k-1$ and $h=1,\ldots,k$. For the case of spherical data ($k=3$) the inverse transformation of \eqref{eq:xth} would be 
\begin{equation}
\label{eq:thx}
\Theta_1=\cos^{-1}(X_1/R)\quad\mbox{and}\quad\Theta_2=\tan^{-1}(X_3/X_2).
\end{equation}
Functions \eqref{eq:thx} are depicted in Figure \ref{fig:coef}. We can see that $X_1$ has an inverse influence on $\Theta_1$, $X_2$ also has an inverse relationship with $\Theta_2$, but discontinuous, i.e., positive $X_2$ values produce larger values of $\Theta_2$ than negative values. Finally $X_3$ has a direct relation with $\Theta_2$. 

Depending on the dimension $k-1$ of the directional response vector $\bTheta$, if $k-1\geq 2$, we can partition $\bTheta=(\bTheta_I, \bTheta_{II})$ and define a (directional+linear)-directional regression model $f(\btheta_I\mid\btheta_{II},\bz)$, as in \eqref{eq:reg} with $f(\btheta)$ replaced by $f(\btheta\mid\bz)$. Here $\bTheta_I$ would be the response variable, $\bTheta_{II}$ the directional and $\bZ$ the linear explanatory variables.

\section{Posterior inference}
\label{sec:post}

Let $\bTheta_1,\bTheta_2,\ldots,\bTheta_n$ be a sample of size $n$ of dimension $(k-1)$ vectors such that $\bTheta_i\mid f\sim f$, independently, and $f\sim\PPT_{k-1}(\alpha,\varphi,f_0,M)$, as in \eqref{eq:PPT}. We consider a data augmentation approach \citep{tanner:91} by defining latent resultant lengths $R_1,R_2,\ldots,R_n$ such that $(\bTheta_i,R_i)$ define the polar coordinate transformation of the multivariate linear vectors $\bX_i=(X_{1,i},\ldots,X_{k,i})$, for $i=1,\ldots,n$.

Then, the likelihood for $\{\bY_{m,j_m}\}$, $\bj_m=(j_{m,1},\ldots,j_{m,k})$, $j_{m,l}=1,\ldots,2^m$, $l=1,\ldots,k$ and $m=0,1,\ldots,M-1$, given the extended data, is 
$$\mbox{lik}(\bY\mid\data)=\prod_{i=1}^n\prod_{m=1}^M Y_{m,\bj_m^{(\bx_i(\btheta_i,r_i))}}=
\prod_{m=1}^M\prod_{j_{m,1}=1}^{2^m}\cdots\prod_{j_{m,k}=1}^{2^m}Y_{m,j_{m,1},\ldots,j_{m,k}}^{N_{m,j_{m,1},\ldots,j_{m,k}}},$$ 
where $N_{m,\bj_m}=\sum_{i=1}^n I\left(\bx_i(\btheta_i,r_i)\in B_{m,\bj_m}\right)$. 

Recalling from Definition \ref{def:mPT} that the prior distribution of the vectors $\bY_{m,\bj_m}$ is Dirichlet with parameter $\balpha_{m,\bj_m}$, and noting that the prior is conjugate with respect to this likelihood, then the posterior distribution for the branching probability vectors is
\begin{equation}
\label{eq:posty}
\bY_{m,\bj_m}\mid\btheta,\br\sim\dir(\balpha_{m,\bj_m}+\bN_{m,\bj_m}).
\end{equation}
where $\bN_{m,\bj_m}=\{N_{m+1,\bj_{m+1}},\,\bj_{m+1}\in\{\times_{l=1}^k(2j_{m,l}-1,2j_{m,l})\}\}$. 

Remember that this posterior depends on an extended version of the data. The latent resultant lengths $R_i$'s have to be sampled from their corresponding posterior predictive distribution which is simply 
\begin{equation}
\label{eq:postr}
f(r_i\mid\bY,\btheta_i)\propto\left\{\prod_{m=1}^M Y_{m,\bj_m^{(\bx_i(\btheta_i,r_i))}}\right\}f_0(\bx_i(\btheta_i,r_i))\,r_i^{k-1},
\end{equation}
for $i=1,\ldots,n$.

There are two parameters that have to be defined a-priori, one is the precision parameter $\alpha$ and the other is the centring measure $f_0$. Instead of fixing them, we suggest to assign hyper prior distributions. Say $\alpha\sim\ga(a_\alpha,b_\alpha)$ and for $f_0$, in the case that this is a multivariate normal with mean $\bmu$ and precision matrix $\bI$, we further take $\mu_l\sim\no(\mu_0,\tau_\mu)$, independently for $l=1,\ldots,k$. Conditional posterior distributions for these parameters are:
\begin{equation}
\label{eq:posta}
f(\alpha\mid\bY)\propto\ga(\alpha\mid a_\alpha,b_\alpha)\prod_{m=0}^{M-1}\prod_{\bj_m}\dir(\by_{m,\bj_m}\mid\balpha_{m,\bj_m})
\end{equation}
and 
\begin{equation}
\label{eq:postmu}
f(\mu_l\mid\btheta,\br)=\no\left(\mu_l\left|\,\frac{n \bar{x}_l(\btheta,\br)+\tau_\mu \mu_0}{n+\tau_\mu},n+\tau_\mu\right.\right),
\end{equation}
where $\bar{x}_l(\btheta,\br)=\frac{1}{n}\sum_{i=1}^n x_{l,i}(\btheta_i,r_i)$, for $l=1,\ldots,k$. 

In the linear-directional regression setting, together with each $\bTheta_i$ we observe a set of linear covariates $\bZ_i$, such that $\bTheta_i\mid\bZ_i,f\sim f(\btheta_i\mid\bz_i)$ for $i=1,\ldots,n$, as in Section \ref{sec:reg}. Augmenting the data, as before, with the latent resultant lengths, we define the one to one transformation of polar coordinates $(\btheta_i,R_i)$ to the multivariate linear vectors $\bepsilon_i$, where $\bepsilon_i=\bX_i-\Gamma\,\bZ_i$ for $i=1,\ldots,n$. Then, conditional posterior distribution of the branching probabilities $\bY_{m,\bj_m}$ is as in \eqref{eq:posty} with $N_{m,\bj_m}=\sum_{i=1}^nI\left(\bepsilon_i(\btheta_i,r_i)\in B_{m,\bj_m}\right)$, where the notation emphasises the dependence on $\btheta_i$ and $r_i$. 

To complete the full Bayesian regression model, we need to specify the prior for the regression coefficients $\Gamma=(\gamma_{l,h})$ for $l=1,\ldots,k$ and $h=1,\ldots,p$. For simplicity we take $\gamma_{l,h}\sim\no(0,\tau_\gamma)$ independently $\forall (l,h)$. Therefore, the conditional posterior distribution for each $\gamma_{l,h}$, given the observed extended data, has the form 
\begin{equation}
\label{eq:postb}
f(\gamma_{l,h}\mid\bY,\btheta,\br)\propto \no(\gamma_{l,h}\mid 0,\tau_\gamma)\prod_{i=1}^n\left\{\prod_{m=1}^M Y_{m,\bj_m^{(\bepsilon_i(\btheta_i,r_i,\Gamma))}}\right\}f_0(\bepsilon_i(\btheta_i,r_i,\Gamma)), 
\end{equation}
where now the notation also emphasises the dependence on $\Gamma$. 

With equations \eqref{eq:posty} -- \eqref{eq:postmu}, plus \eqref{eq:postb} (in the case of linear-directional regression), we can implement a Gibbs sampler \citep{smith&roberts:93}. Sampling from \eqref{eq:posty} and \eqref{eq:postmu} is straightforward, however, to sample from \eqref{eq:postr}, \eqref{eq:posta} and \eqref{eq:postb} we will require MH steps \citep{tierney:94}. In those cases we suggest random walk proposal distributions with the following specifications. Dropping the indexes, at iteration $(t+1)$: 
sample $r^*$ from $\ga(\kappa_r,\kappa_r/r^{(t)})$ and accept it with probability $$\pi(r^*,r^{(t)})=\frac{f(r^*\mid\bY,\btheta)\ga(r^{(t)}\mid\kappa_r,\kappa_r/r^*)}{f(r^{(t)}\mid\bY,\btheta)\ga(r^*\mid\kappa_r,\kappa_r/r^{(t)})};$$
sample $\alpha^*$ from $\ga(\kappa_\alpha,\kappa_\alpha/\alpha^{(t)})$ and accept it with probability $$\pi(\alpha^*,\alpha^{(t)})=\frac{f(\alpha^*\mid\bY)\ga(\alpha^{(t)}\mid\kappa_\alpha,\kappa_\alpha/\alpha^*)}{f(\alpha^{(t)}\mid\bY)\ga(\alpha^*\mid\kappa_\alpha,\kappa_\alpha/\alpha^{(t)})};$$
and sample $\gamma^*$ from $\no(\gamma^{(t)},\kappa_\gamma)$ and accept it with probability $$\pi(\gamma^*,\gamma^{(t)})=\frac{f(\gamma^*\mid\bY,\btheta,\br)\no(\gamma^{(t)}\mid \gamma^*,\kappa_\gamma)}{f(\gamma^{(t)}\mid\bY,\btheta,\br)\no(\gamma^*\mid \gamma^{(t)},\kappa_\gamma)}.$$
In any case the acceptance probability $\pi$ is truncated to the interval $[0,1]$. 

\begin{figure}
\centerline{\includegraphics[scale=0.45]{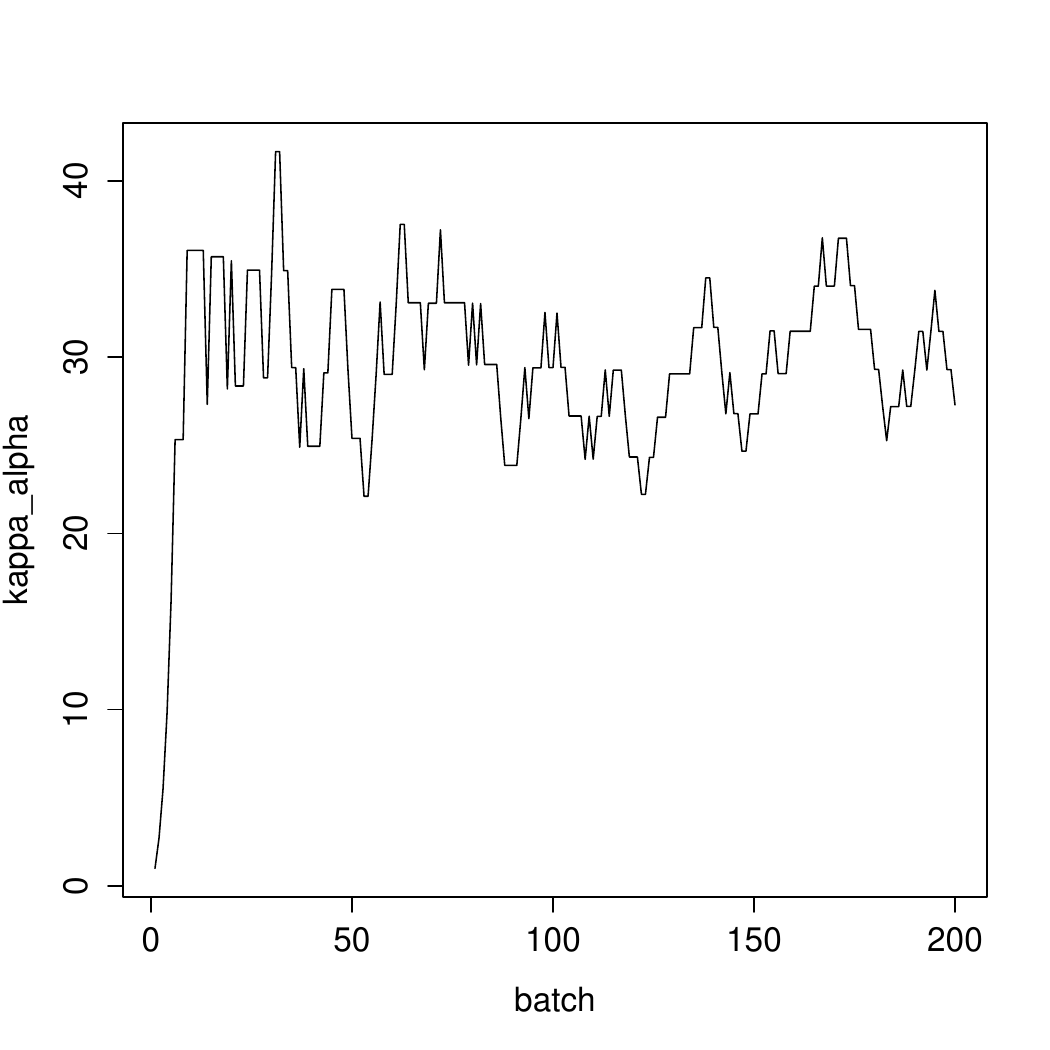}
\includegraphics[scale=0.45]{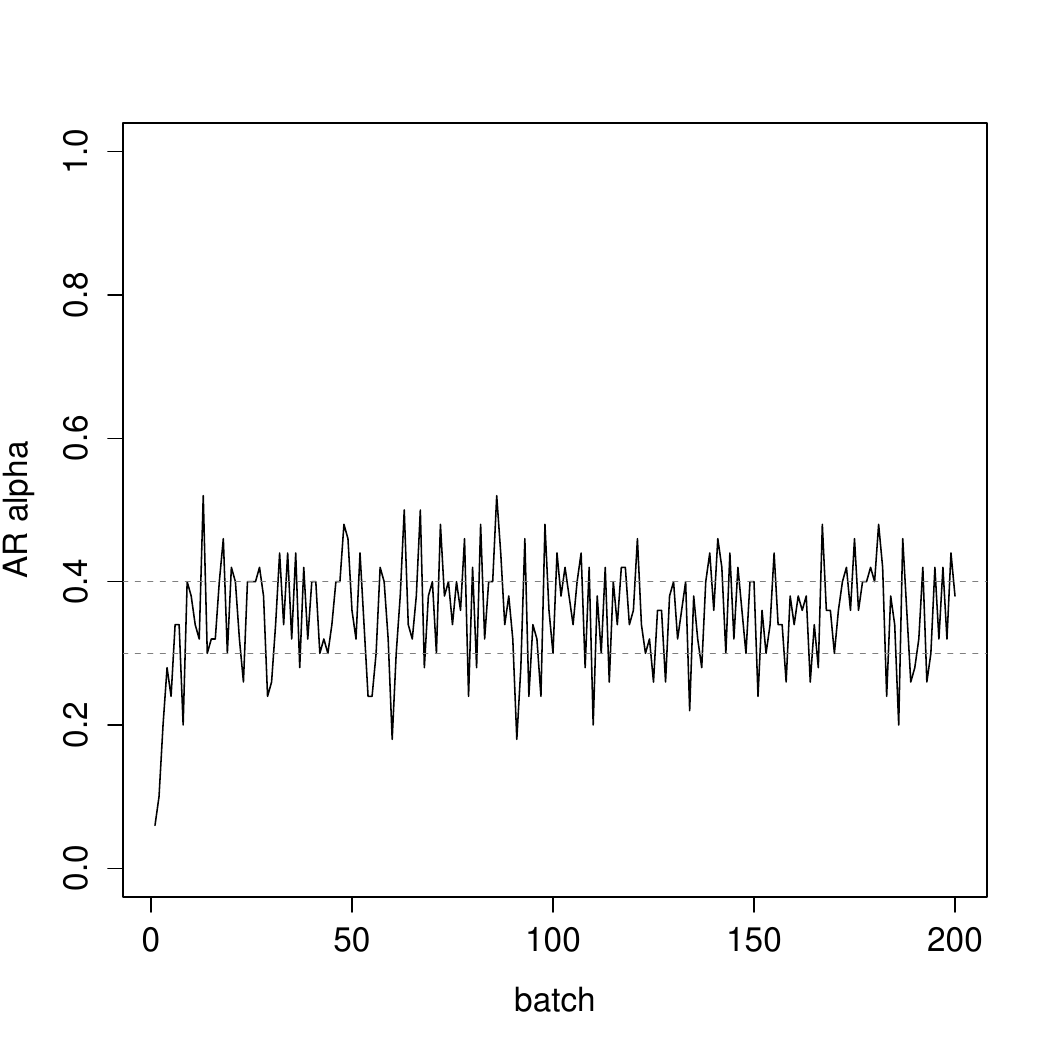}}
\caption{{\small Performance of MH adaptation algorithm for $\alpha$ in dataset B15. $\kappa_\alpha$ tuning parameter (left) and acceptance rate (right).}}
\label{fig:realB19ak}
\end{figure}

The parameters $\kappa_r$, $\kappa_\alpha$ and $\kappa_\gamma$ are nonnegative tuning parameters that determine the acceptance probabilities. As suggested by \cite{roberts&rosenthal:09}, dropping the subindex, $\kappa$ parameters can be adapted every certain amount of iterations, inside the MCMC algorithm, to achieve a target acceptance rate. Differing slightly from the proposal in \cite{roberts&rosenthal:09}, instead of considering a single target acceptance rate, we consider the interval $[0.3, 0.4]$ which, according to \cite{robert&casella:10}, define optimal acceptance rates in random walk MH steps. Specifically, our adaptation method uses batches of 50 iterations and for every batch $b$, we compute the acceptance rate $AR^{(b)}$ and increase $\kappa^{(b+1)}$ to $\kappa^{(b)}e^{b^{1/2}}$ if $AR^{(b)}<0.3$ and decrease $\kappa^{(b+1)}$ to $\kappa^{(b)}e^{-b^{1/2}}$ if $AR^{(b)}>0.4$. For the examples considered here we use $\kappa^{(1)}=1$ as starting value for all MH steps. Figure \ref{fig:realB19ak} shows the performance of this adaptive algorithm for the precision parameter $\alpha$ in the analysis of dataset B15 (Section \ref{ssec:real}). The left panel shows the values of tuning parameter $\kappa_\alpha$ that stabilise around the value of 30. The right panel shows that the acceptance rate is kept around the target interval $[0.3,0.4]$ as desired. 

As a measure of goodness of fit, for each prior scenario we will compute the logarithm of the pseudo marginal likelihood (LPML), originally suggested by \cite{geisser&eddy:79} and the deviance information criteria (DIC) introduced by \cite{spiegelhalter&al:02}. In particular, we will choose the best value for the hyper-parameters by comparing these measures. 

In general, the projected P\'olya tree directional model $f(\btheta)$, as in \eqref{eq:PPT} or \eqref{eq:PPTn}, does not have a closed analytic expression. It depends on an integral with respect to the latent $R$. We propose to approximate the density via numerical quadrature. Say, if $0=r^{(0)}<r^{(1)}<\cdots<r^{(T)}<\infty$ is a partition of the positive real line, then \eqref{eq:PPT} is approximated by
$$f(\btheta)\approx \sum_{t=1}^T\left\{\prod_{m=1}^M Y_{m,\bj_m^{(\bx(\btheta,r^{(t)}))}}\right\}2^{kM} f_0(\bx(\btheta,r^{(t)}))(r^{(t)})^{k-1}\left\{\prod_{l=1}^{k-2}(\sin\theta_l)^{k-l-1}\right\}(r^{(t)}-r^{(t-1)}).$$
Specifically we take $T=100$, $r^{(T)}=||\bmu||+4$ and the rest $r^{(t)}$ equally spread. The code that implements the model and the data are available as Supplementary Material.

\section{Numerical studies}
\label{sec:numerical}

\subsection{Spherical data}

There are two cases where directional data can be easily understood and visualise, circular data ($k=2$) and spherical data ($k=3$). In the examples below we concentrate on the latter scenario. For this, let us envision a sphere and imagine an arrow piercing the sphere in the center from south to north, and a table cutting the sphere right in the equator. Consider an arbitrary point on the surface of the sphere. The angle $\theta_1$, called \emph{colatitud}, is the angle from the arrow to the point, measured from north to south. The angle $\theta_2$, called \emph{longitude}, is the angle measured anti clock-wise of the projection of the point to the table. The pair $(\theta_1,\theta_2)$ characterises any point on the sphere surface. 

Here we present four numerical studies. In the first study we sample prior paths from the projected P\'olya tree centred around the projected normal. The objective is to understand the role of the parameters $\bmu$ in the joint and marginal directional distributions of $(\Theta_1,\Theta_2)$ as well as to characterise the mean directions $\nu_l$, concentrations $\varrho_l$, for $l=1,2$, and correlation measures induced. We also illustrate the shape of the regression curves (conditional means) induced. 

The second study assesses the performance of our model under a control scenario. We sample from a projected normal and a mixture of projected normals to validate our posterior inference procedure and compare with the ground truth. 

The third study performs Bayesian inference for two particular datasets. The objective is to show the flexibility of our model to capture multimodalities in the joint and marginal distributions of $(\Theta_1,\Theta_2)$, as well as to characterise nonlinear dependencies and the regression curves. 

The fourth study illustrates an analysis of spherical data with linear covariates. Here we are able to illustrate the directional-linear regression as well as the directional-directional+linear regression. 

\subsection{Simulated prior paths}

We first illustrate some paths of the projected P\'olya tree centred around the projected normal, as in \eqref{eq:PPTn}. We took three levels of the partition $M=3$, a precision parameter $\alpha=1$, a function $\varphi(m)=m^{\delta}$ with $\delta=1.1$, and different values of $\bmu$. For each setting we sampled ten paths (densities) from the model. The joint and marginal densities of $\Theta_l$, $l=1,2$, were approximated numerically with $T=100$ points.

\begin{figure}
\centerline{\includegraphics[scale=0.31]{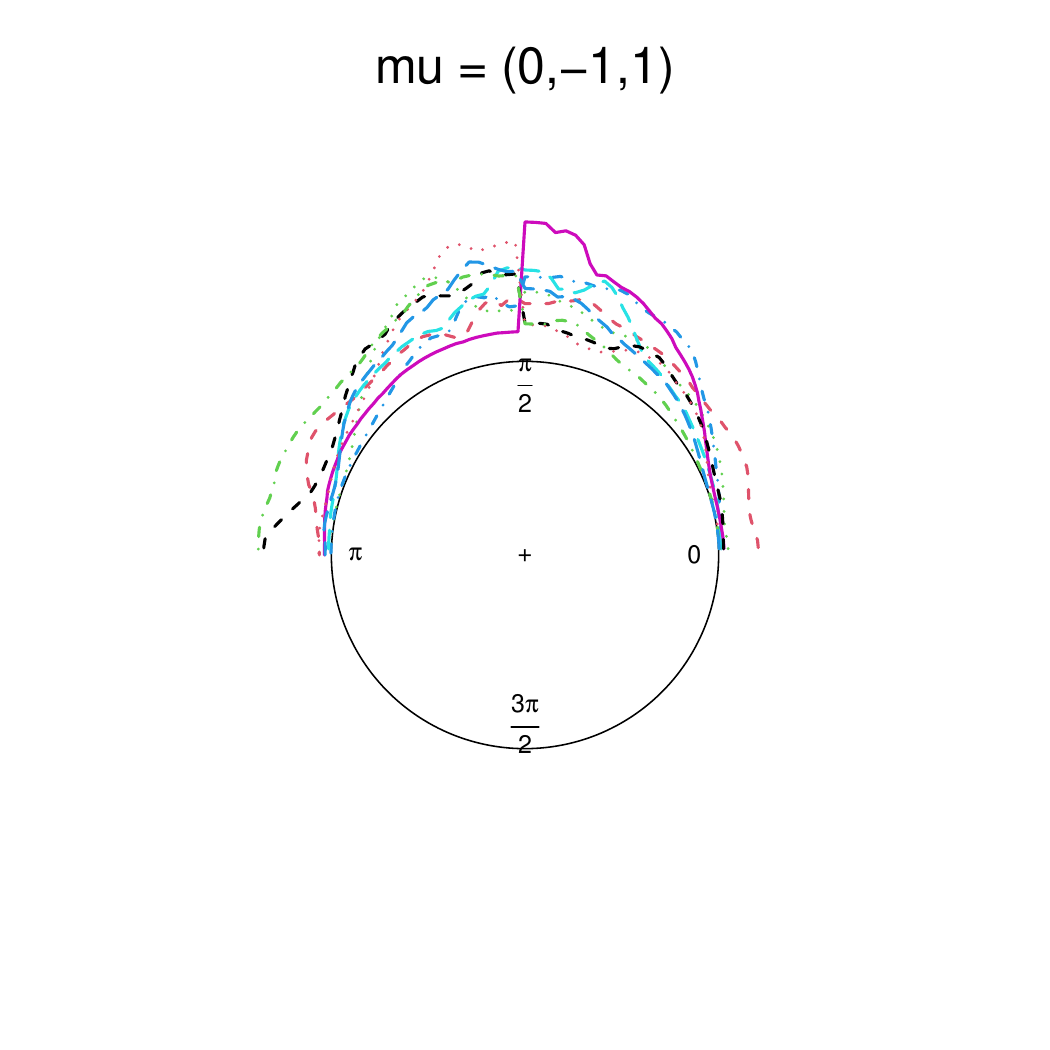}
\includegraphics[scale=0.31]{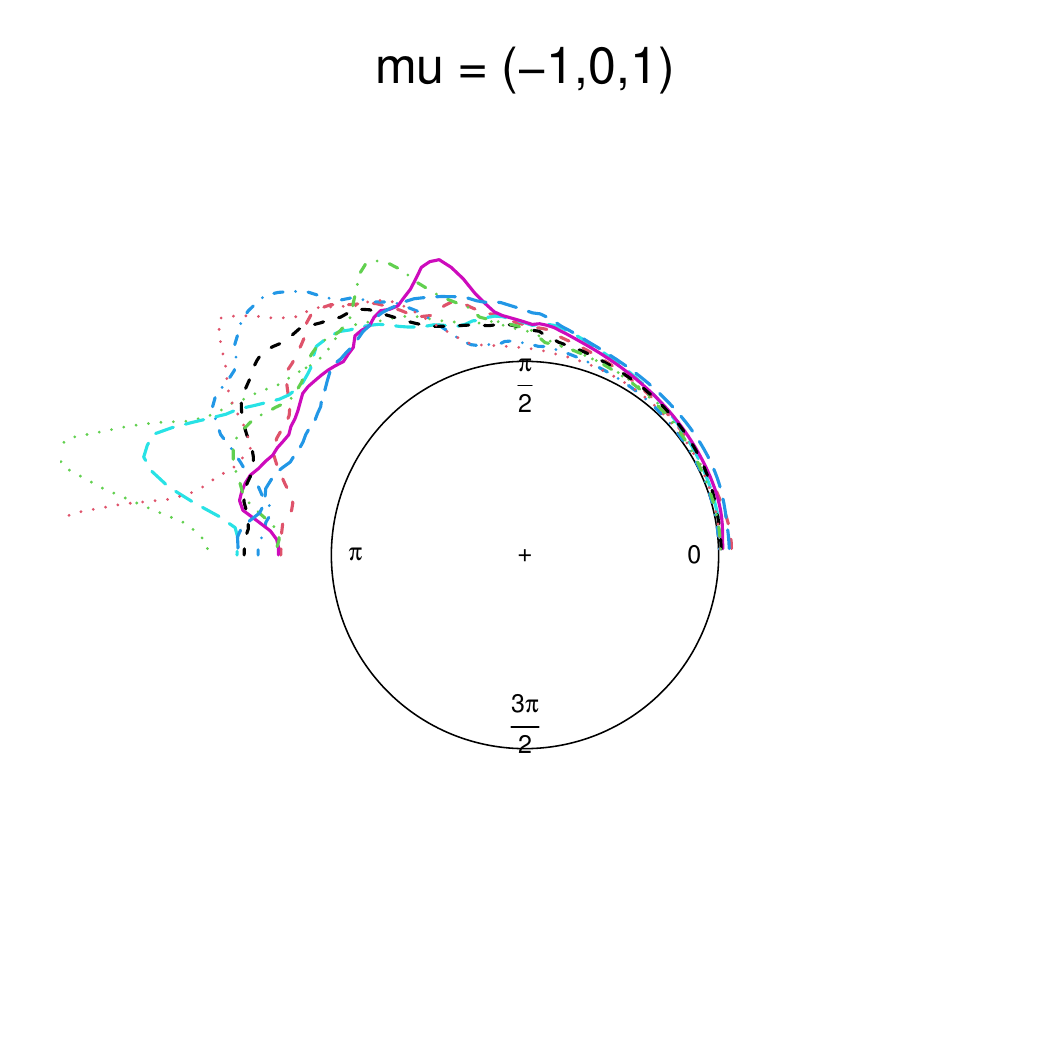}
\includegraphics[scale=0.31]{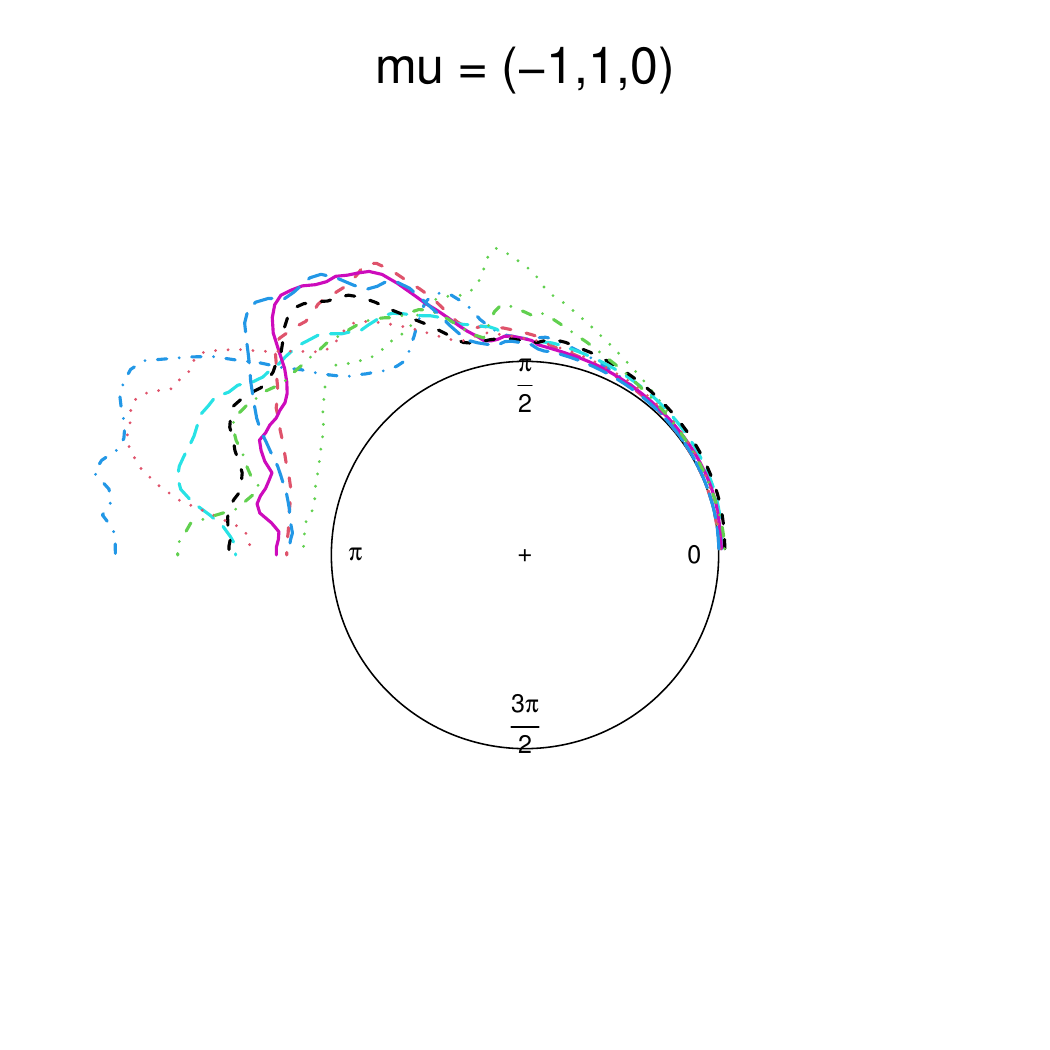}}
\centerline{\includegraphics[scale=0.31]{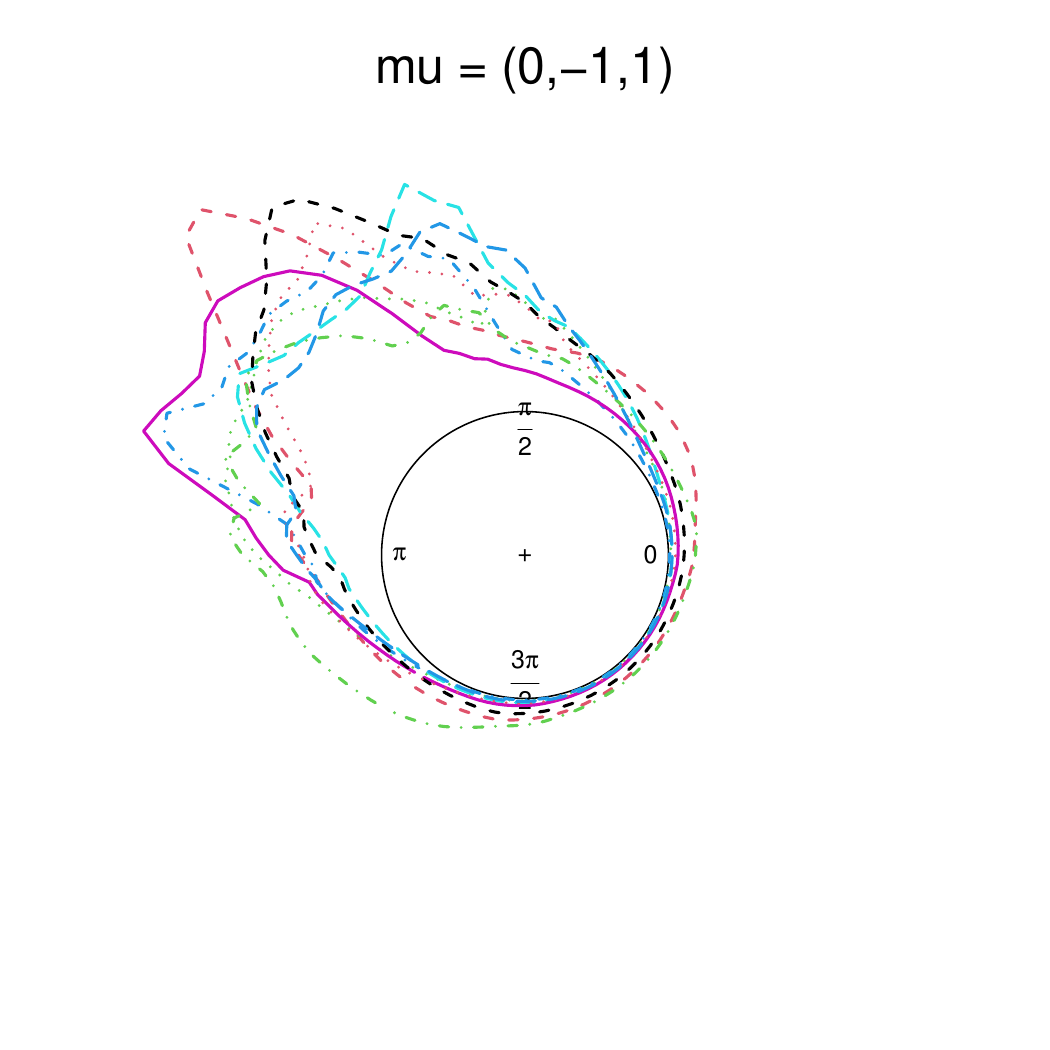}
\includegraphics[scale=0.31]{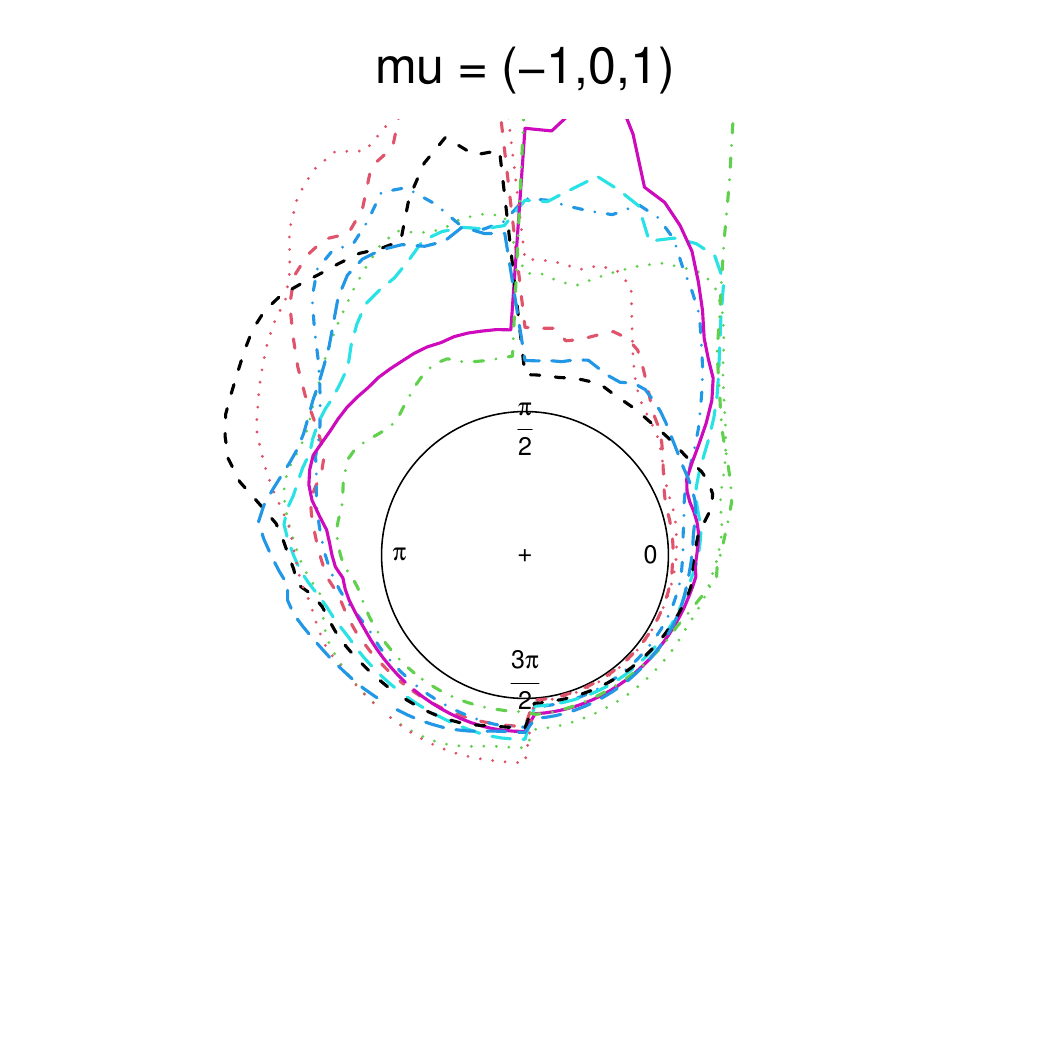}
\includegraphics[scale=0.31]{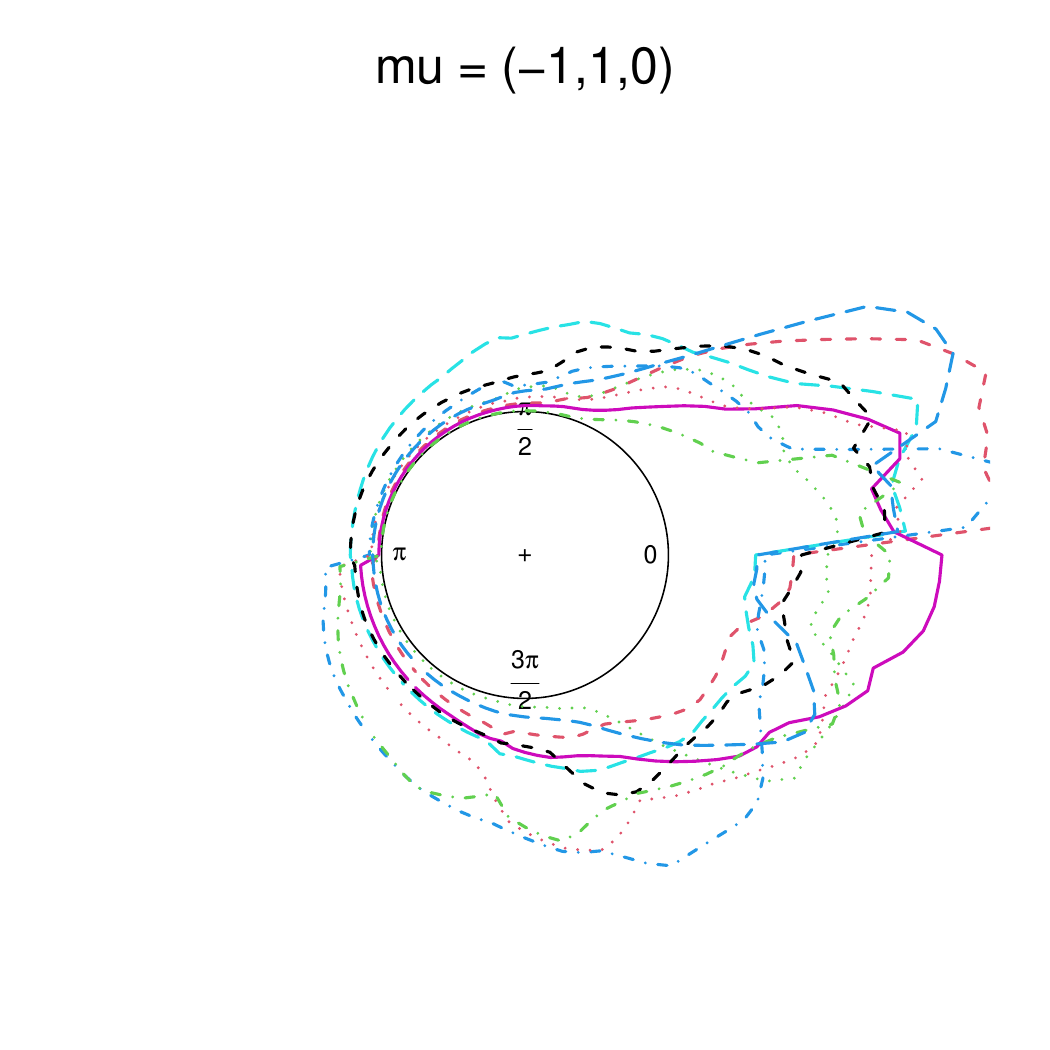}}
\centerline{\includegraphics[scale=0.31]{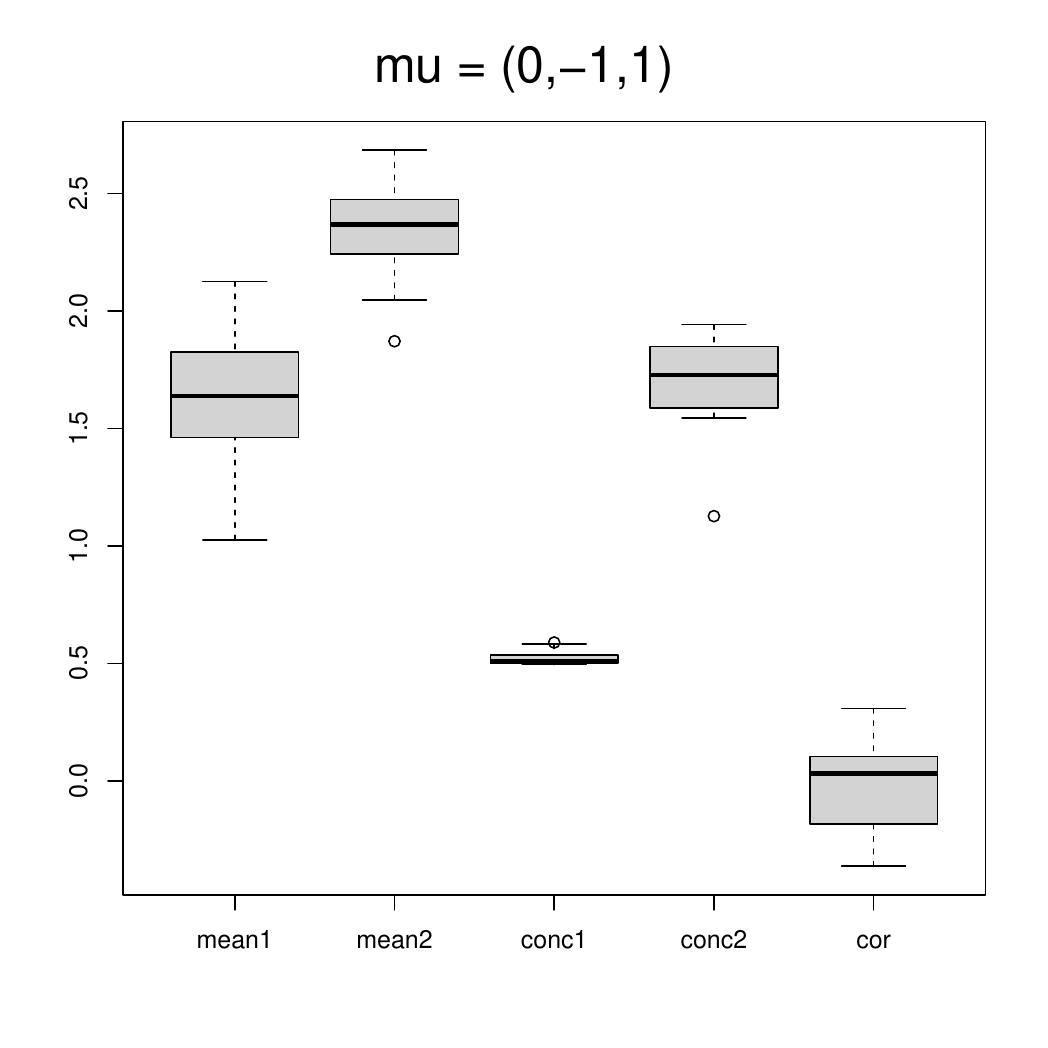}
\includegraphics[scale=0.31]{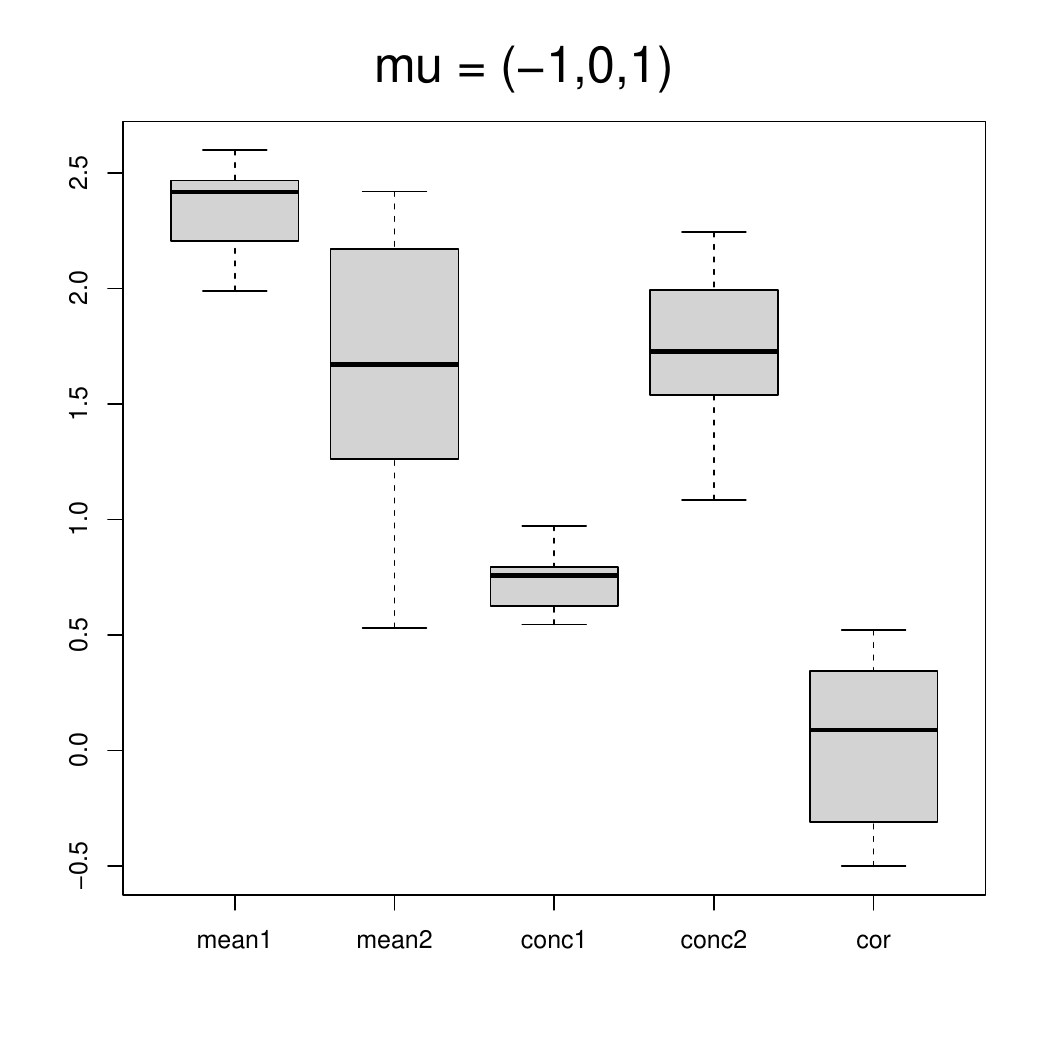}
\includegraphics[scale=0.31]{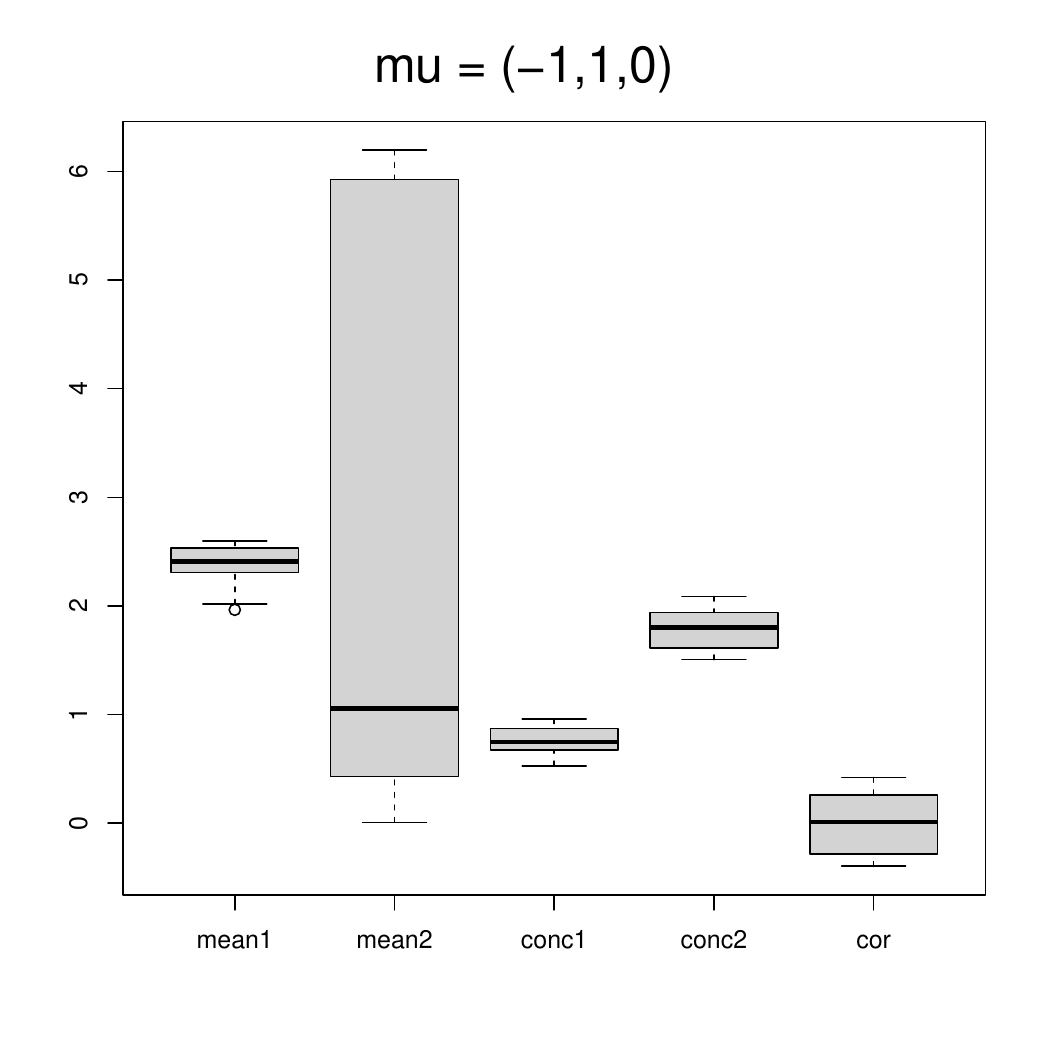}}
\caption{{\small Simulated densities of the prior $\PPT_2$ with $M=3$, $\alpha=1$, $\delta=1.1$. For varying $\bmu'$ (across columns). Marginal densities of $\Theta_1$, $\Theta_2$ and moments (across rows).}}
\label{fig:prior2}
\end{figure}

Figure \ref{fig:prior2} contains nine panels which correspond to $\bmu'=(0,-1,1)$ (first column), $\bmu'=(-1,0,1)$ (middle column) and $\bmu'=(-1,1,0)$ (third column). Marginal densities for $\Theta_1$ and $\Theta_2$ are shown around the circle in the first and second rows, respectively. For $\Theta_1$ we note that the support is bounded to $[0,\pi]$, half upper part of the circle, and that the densities are not periodic, $f(0)\neq f(\pi)$; whereas for $\Theta_2$, the support is $[0,2\pi]$, the densities go around the entire circle and are periodic, $f(0)=f(2\pi)$. If for $\Theta_1$ we completed the plot up to $2\pi$, the density in $[\pi,2\pi]$ would be a reflection of the density in $[0,\pi]$. Third row shows some population moments as boxplots, the mean marginal directions, $\nu_1$ and $\nu_2$, marginal concentrations, $\varrho_1$ and $\varrho_2$, given in \eqref{eq:moments}, and the correlation measure $\rho$ given in \eqref{eq:cor}. 

When we take $\bmu'=(0,-1,1)$, paths for $\Theta_1$ and $\Theta_2$ seem to have a predominant mode around $\pi/2$ and $3\pi/4$, respectively, with the paths for $\Theta_1$ more concentrated than those for $\Theta_2$. For $\bmu'=(-1,0,1)$ our projected P\'olya tree shows more variability in both $\Theta_1$ and $\Theta_2$ and their correlation also shows a large variability covering values between $-1$ and $1$. When we take $\bmu'=(-1,1,0)$ paths for $\Theta_1$ and $\Theta_2$ have almost opposite modes with directions $3\pi/4$ and $0$, respectively. The mean marginal direction $\nu_2$ shows a huge dispersion in the boxplot, which can be interpreted as very diverse locations along the whole interval $[0,2\pi]$. 

\begin{figure}
\centerline{\includegraphics[scale=0.31,angle=270,origin=c]{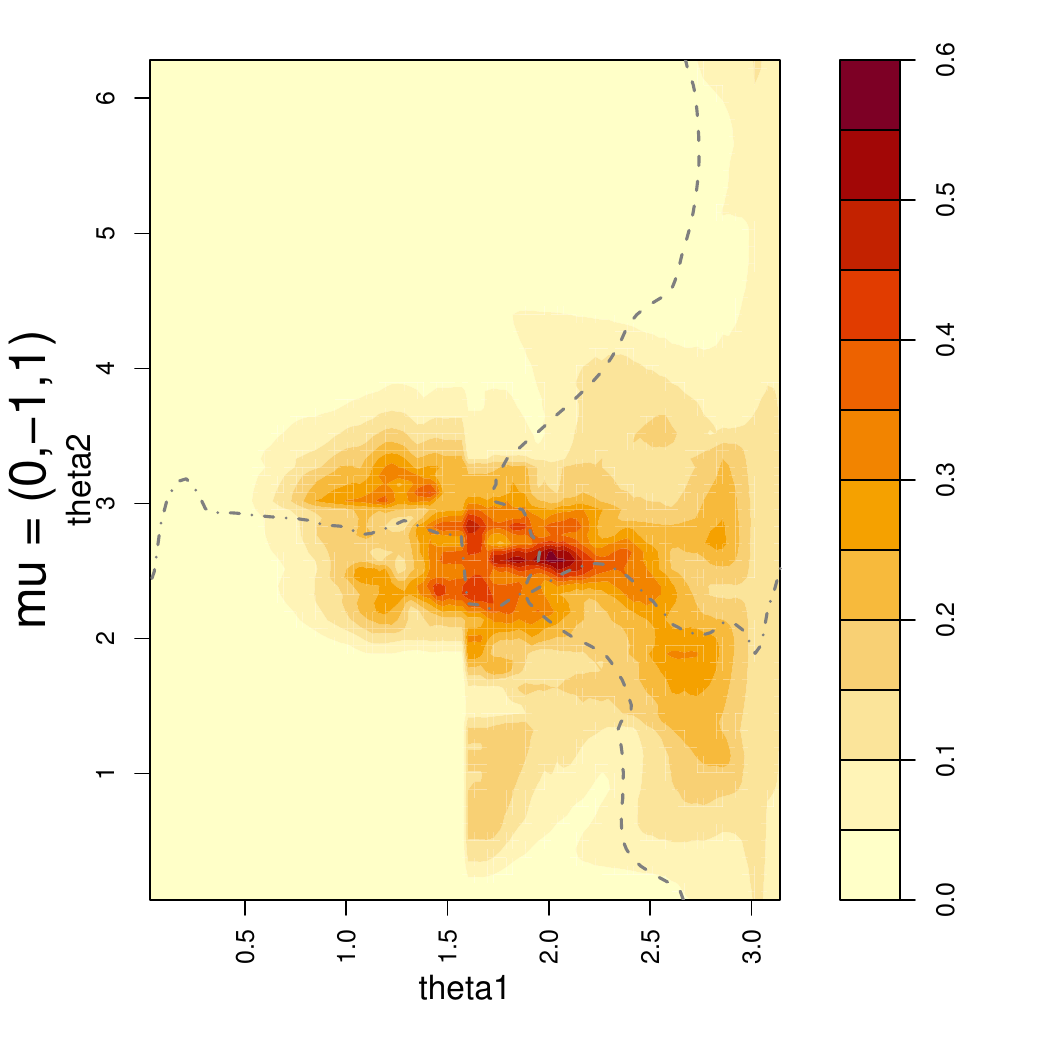}
\includegraphics[scale=0.31,angle=270,origin=c]{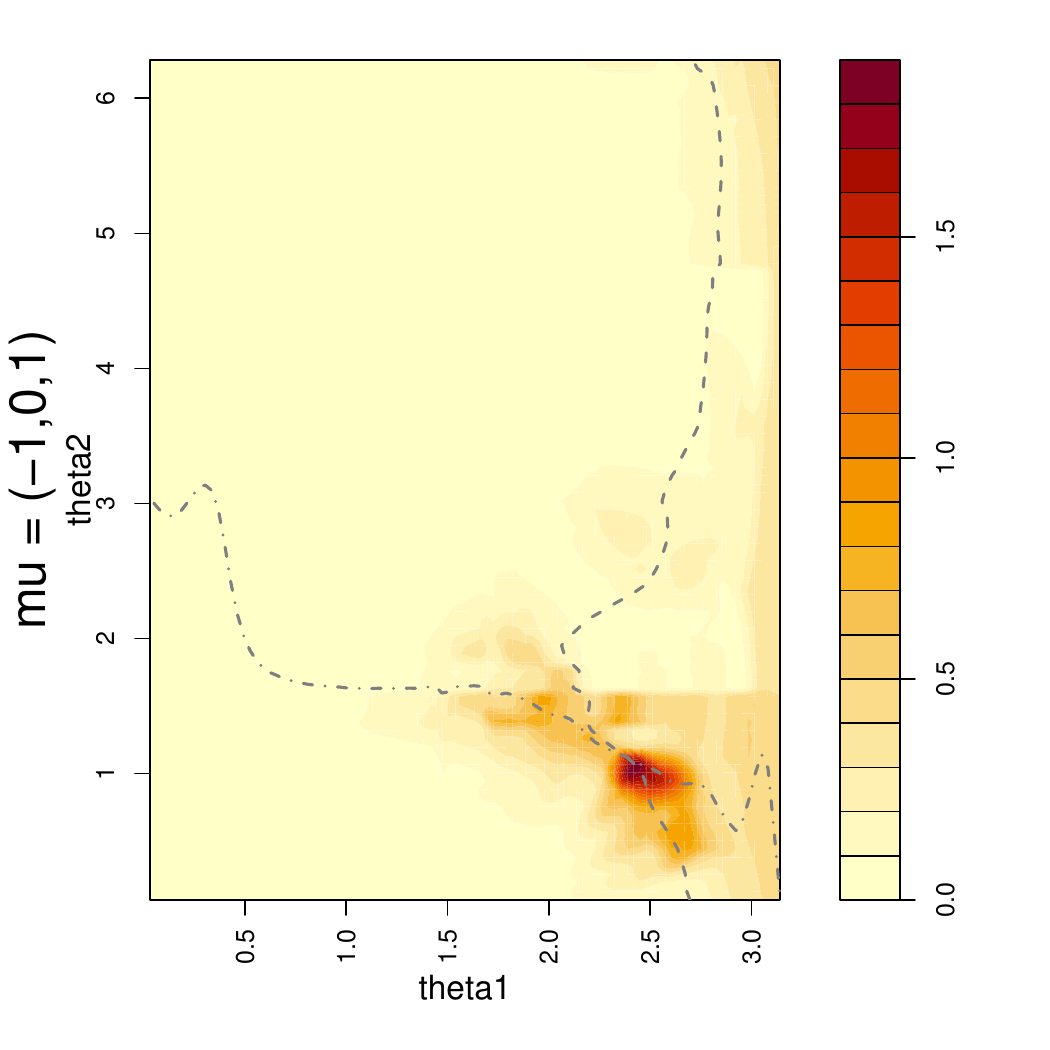}
\includegraphics[scale=0.31,angle=270,origin=c]{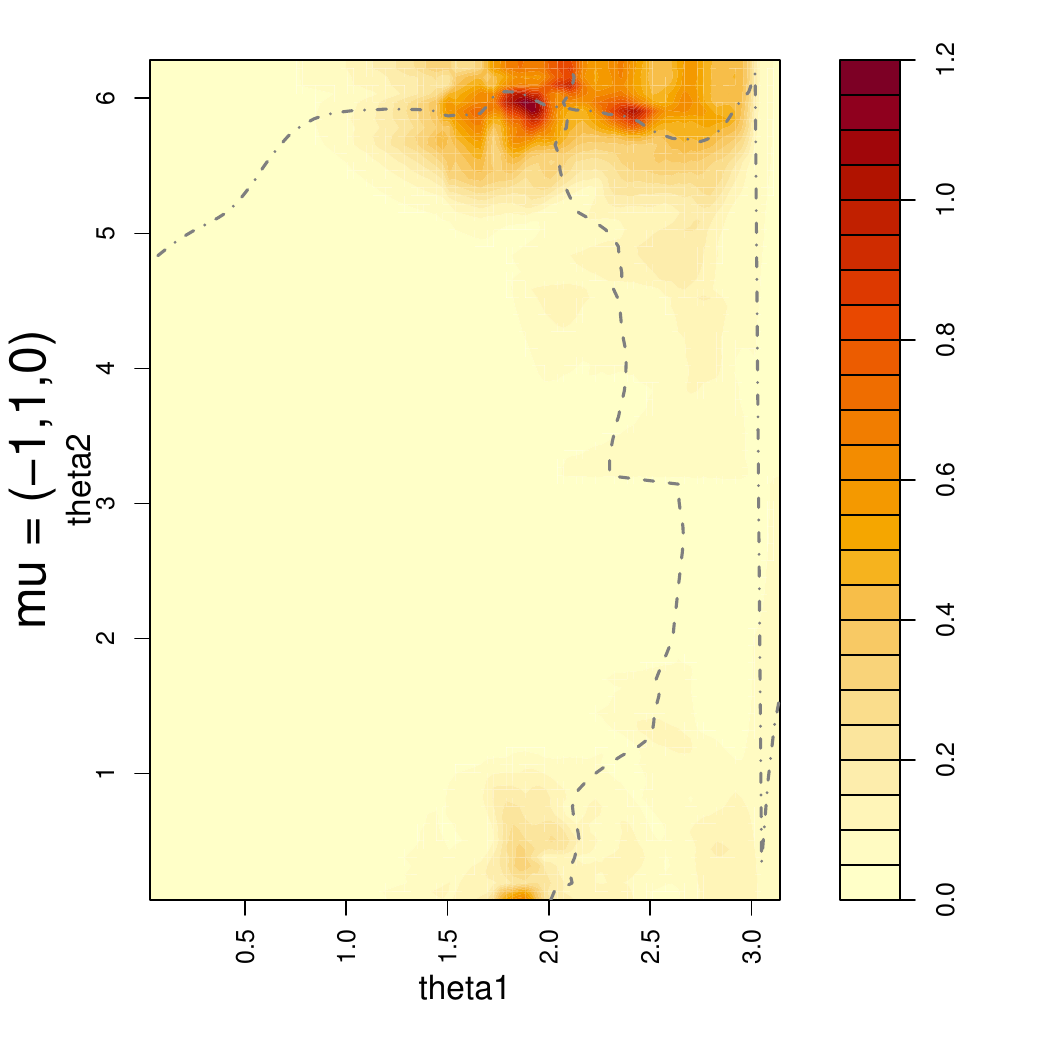}}
\caption{{\small Simulated densities of the prior $\PPT_2$ with $M=3$, $\alpha=1$, $\delta=1.1$, for varying $\bmu'$. Joint density (contour plot) and conditional means $\E(\Theta_1\mid\theta_2)$ (dashed line) and $\E(\Theta_2\mid\theta_1)$ (dashdotted line).}}
\label{fig:priorC12}
\end{figure}

To illustrate the shapes of the joint distributions, Figure \ref{fig:priorC12} contains single simulations for each of the three different values of $\bmu$ considered in Figure \ref{fig:prior2}. Contours of the joint densities are shown as heat maps with darker colour indicating higher density. The first two cases have negative dependence and the last case has a positive dependence. As for a circular variable, the best way of plotting a density is on a circle, for a bivariate directional vector, the best plot would be a cylinder, rolling back top and bottom edges of the plots. By doing so, contour for $\bmu'=(-1,1,0)$  that could be seen having two separate clouds, would turn into a single cloud. Additionally in the three panels, we also include the conditional means $\E(\Theta_1\mid\theta_2)$ and $\E(\Theta_2\mid\theta_1)$, which are functions of the conditioning variable $\theta_2$ and $\theta_1$, respectively. This conditional means can be interpreted as the regression curves. For these case conditional means show diagonal lines, specially in the region with high probability. 

\subsection{Simulated data analysis}
\label{ssec:sim}

\begin{figure}
\centerline{\includegraphics[scale=0.31,angle=270,origin=c]{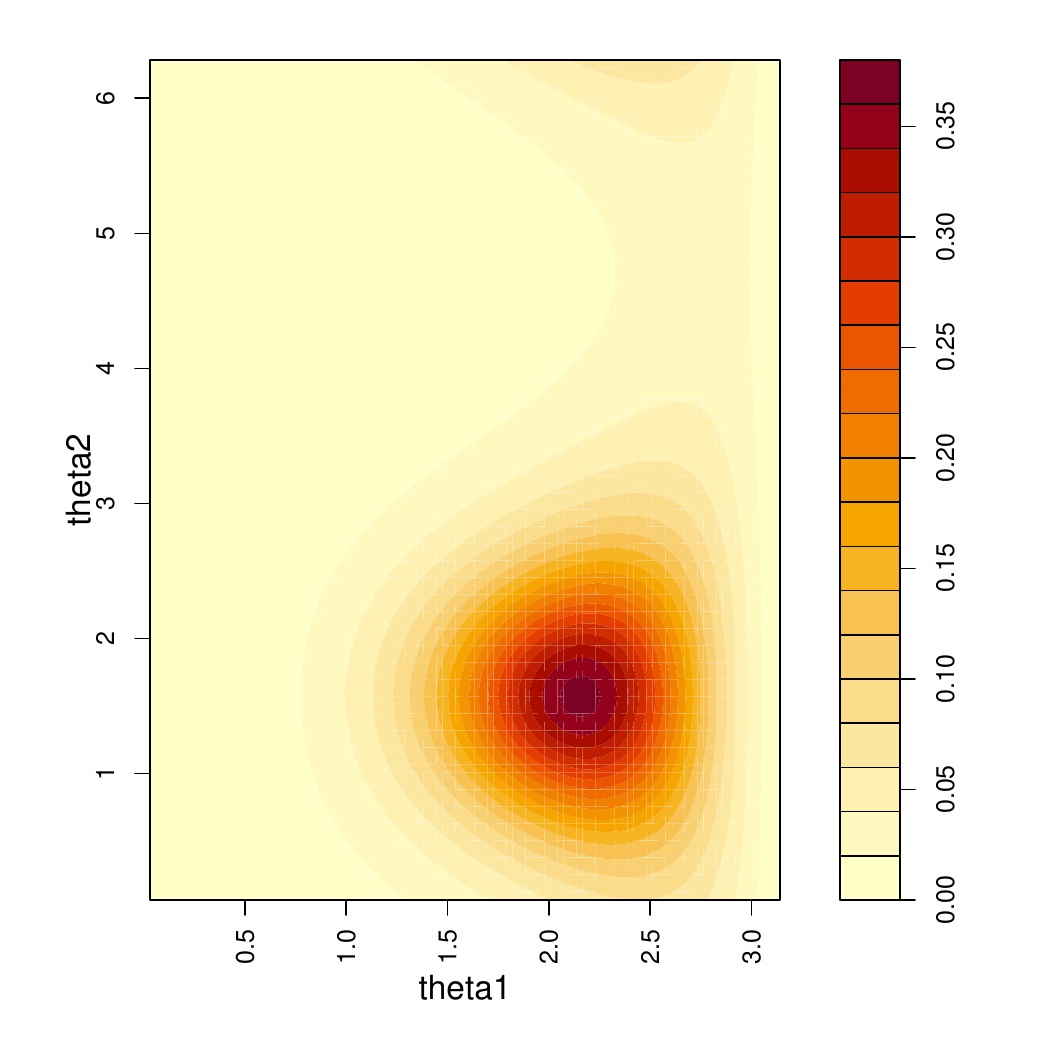}
\includegraphics[scale=0.31,angle=270,origin=c]{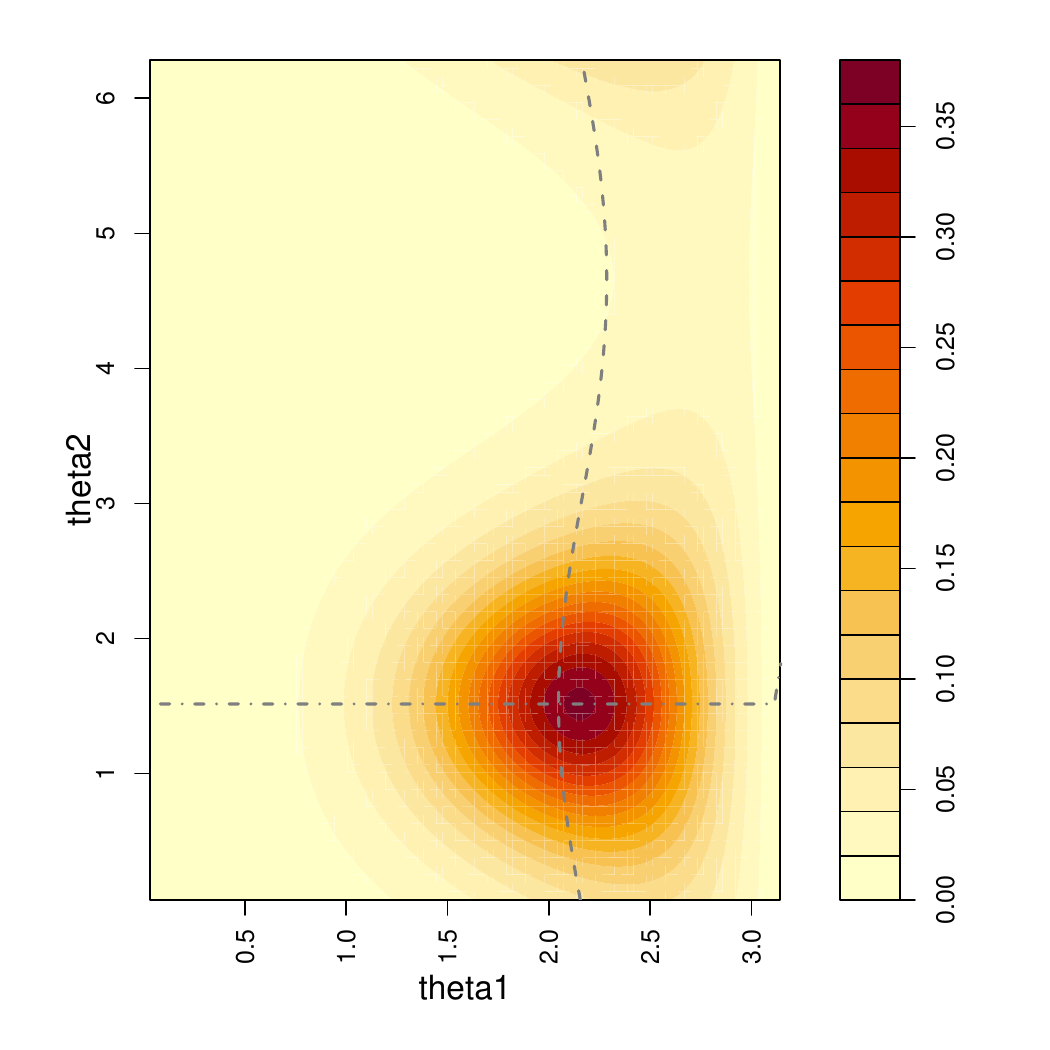}
\includegraphics[scale=0.31,angle=270,origin=c]{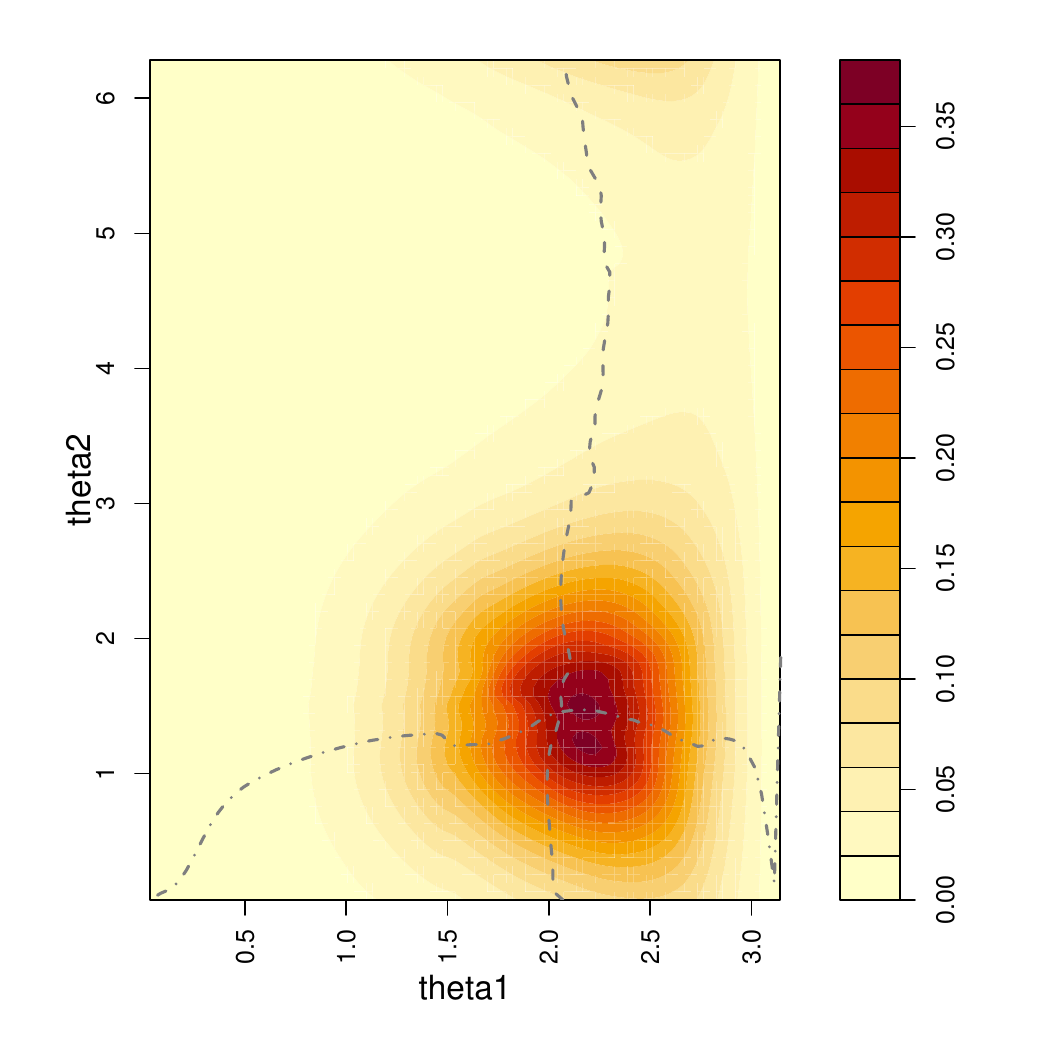}}
\centerline{\includegraphics[scale=0.31,angle=270,origin=c]{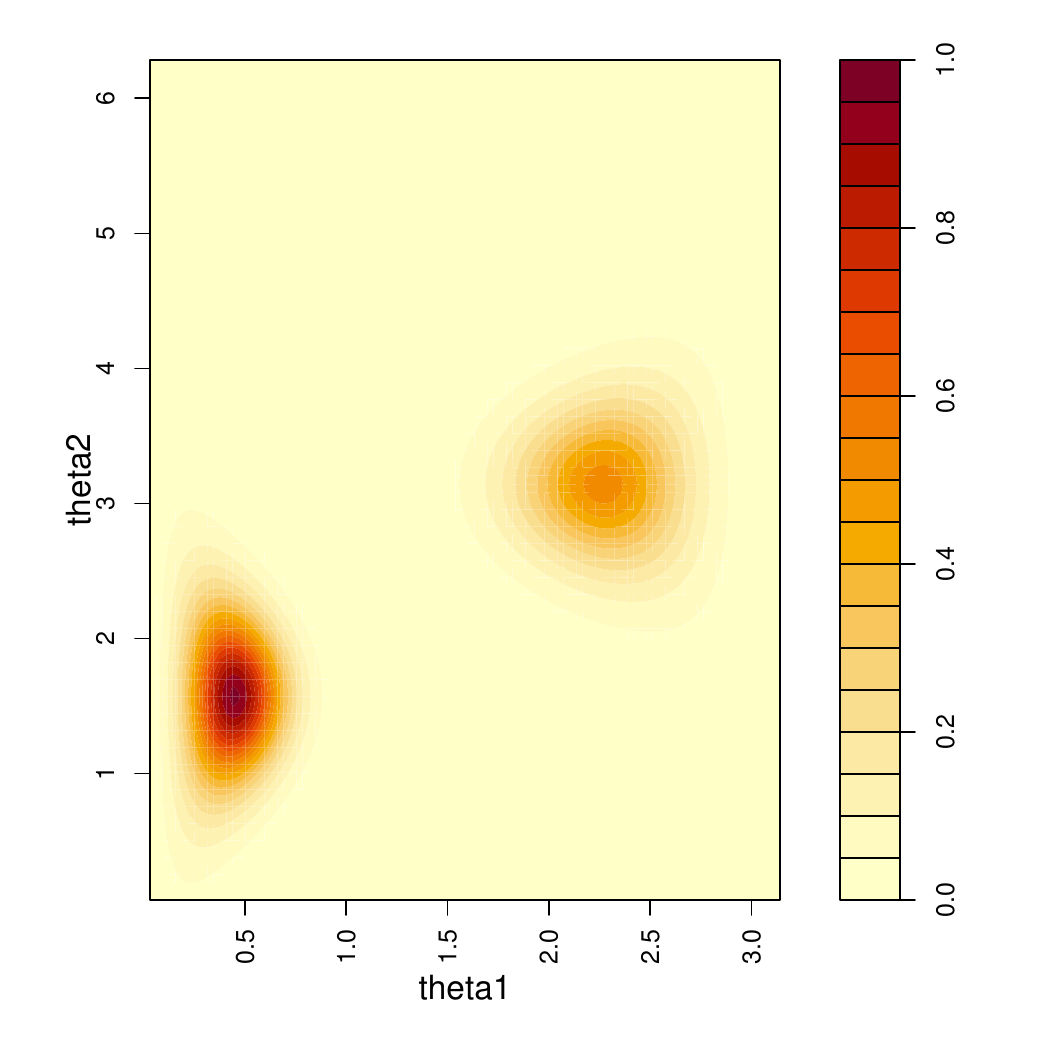}
\includegraphics[scale=0.31,angle=270,origin=c]{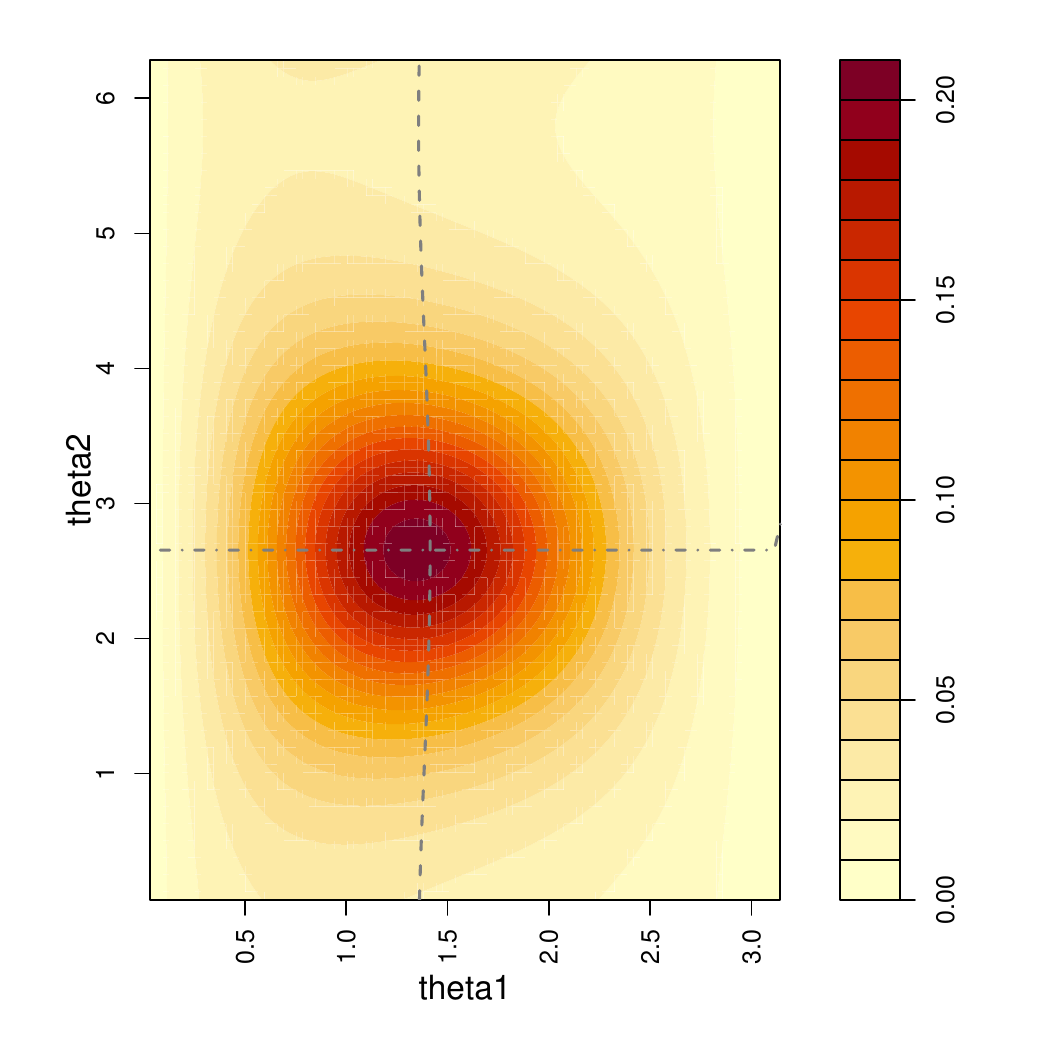}
\includegraphics[scale=0.31,angle=270,origin=c]{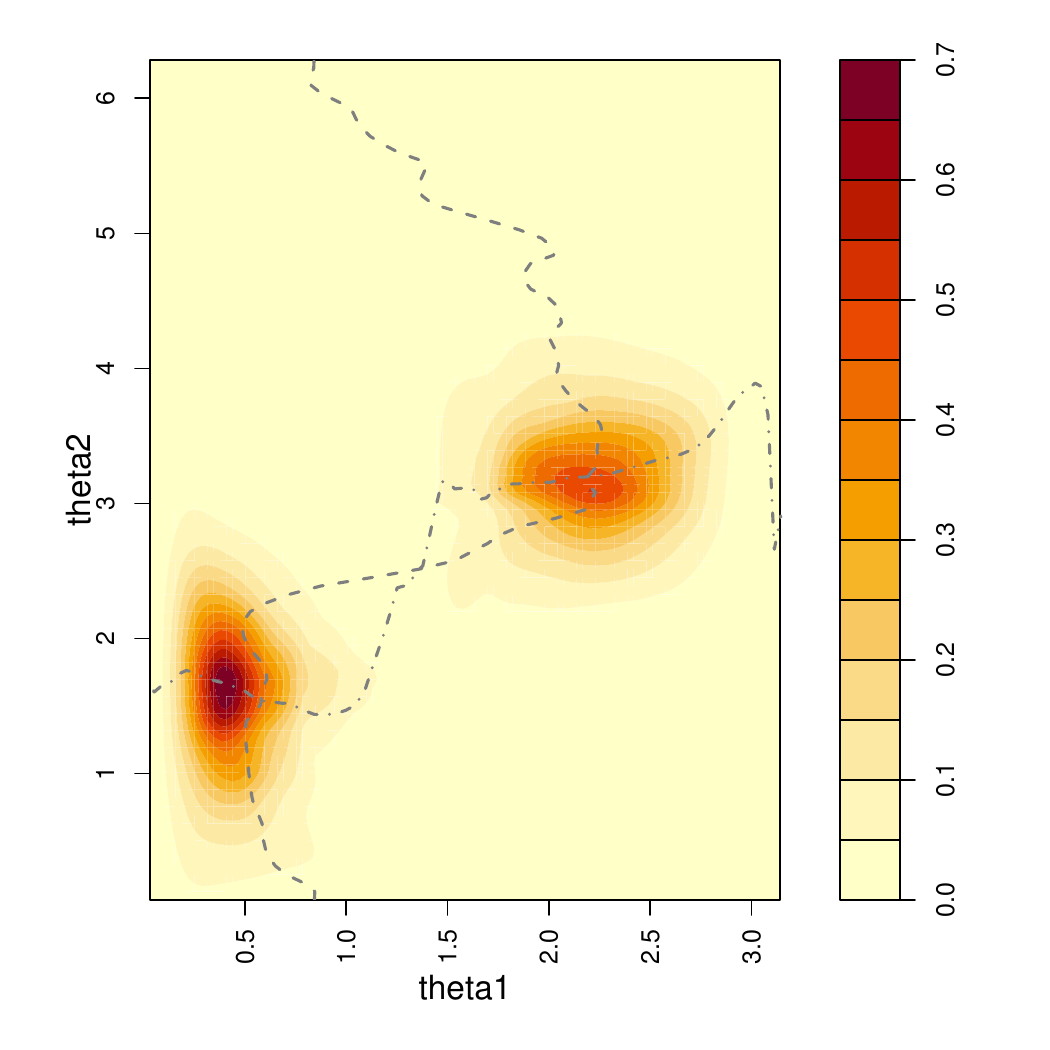}}
\caption{{\small Contour plots of joint density and conditional means $\E(\Theta_1\mid\theta_2)$ (dashed line) and $\E(\Theta_2\mid\theta_1)$ (dashdotted line). Scenario 1 (first row) and scenario 2 (second row). True (first column), projected normal (second column) and PPT model (third column).}}
\label{fig:sim}
\end{figure}

To assess the performance of our model in a controlled setting, we undertake a simulation study for spherical data with two scenarios. In scenario S1 we sample from a multivariate normal $\bX_i\sim\no_3(\bmu,\bI)$ for $i=1,\ldots,n$ and $\bmu'=(-1,0,1)$. In scenario S2 we sample from a mixture of two multivariate normals of the form $X_i\sim\omega\no_3(\bmu_1,\bI)+(1-\omega)\no_3(\mu_2,\bI)$ for $i=1,\ldots,n$ with $\bmu_1'=(-2,-2,0)$, $\bmu_2'=(5,0,2)$ and $\omega=0.5$. In both cases we transform the $\bX_i$ variables to polar coordinates $(\Theta_{1i},\Theta_{2i})$ using \eqref{eq:thx}, for $i=1,\ldots,n$. In scenario S1 we have a unimodal distribution, whereas in scenario S2 we have a bimodal distribution, as shown in Figure \ref{fig:sim} (first column). We took two sample sizes, $n=100$ and $n=300$ to compare.

Our projected P\'olya tree model ($\PPT$) of Section \ref{sec:model} is centred in a projected normal with the following prior specifications: $M=3$, $\delta=1.1$, a prior for the precision $\alpha\sim\ga(1,2)$ and $\mu_l\sim\no(0,1)$ independent for $l=1,2,3$. Posterior inference was based on our adaptive MCMC with 10,000 iterations with a burn-in of 2,000 and keeping one of every 10$^{th}$ iteration after burn-in. 

For comparison purposes we also fitted a projected normal model and computed two goodness of fit (gof) statistics, the LPML and the DIC. These numbers are reported in Table \ref{tab:sim}. For scenario S1, the best model is the projected normal, as it should be, but the gof statistics do not have a huge difference with respect to those obtained with the $\PPT$. However, for scenario the gof statistics highly support the $\PPT$. In Figure \ref{fig:sim} we present the fitting of the two models for $n=300$. The projected normal (bottom middle panel) behaves terribly for scenario S2, whereas the $\PPT$ (third column) captures the true shapes in both scenarios S1 and S2. Estimated conditional means for S2 are somehow chaotic due to the bimodality.

\subsection{Real data analysis}
\label{ssec:real}

The book \cite{fisher&al:87} presents a series of spherical datasets. We have selected two of them to illustrate the performance of our model. Specifically we have chosen the sets B15 and B19. The first dataset deals with the orientation of the dendritic field at various sites in the retina of 6 cats, in response to different visual stimuli. Dendritic fields are fields with branched forms resembling trees.  The spherical data are defined by latitude ($\vartheta_1$) and longitude ($\vartheta_2$), which take values in $[90,180]$ and $[0,360]$ degrees, respectively. The second dataset deals with orientation of beddings of F$_1$ folds in Ordovician turbidites. In plain words, ``F$_1$'' is a category used by geologists to define the first folding, ``Ordovician'' is the second period of the Paleozoic era, and a ``tubidite'' is the sediment left by an underwater current. The spherical data in this context is defined by two measures, dip ($\vartheta_1$) and dip direction ($\vartheta_2$), which take values in $[0,90]$ and $[0,360]$ degrees, respectively. In both cases, we transform the data into radians by defining $\theta_1=\pi\eta_1/180$ and $\theta_2=2\pi\eta_2/360$. 

In either case we have a sample $(\Theta_{1,i},\Theta_{2,i})$ for $i=1,\ldots,n$, where the sample sizes are 94 and 64, respectively for each datasets. To analyse the data we use the directional (spherical) projected P\'olya tree model ($\PPT$) of Section \ref{sec:model} centred in a projected normal with the following prior specifications: The depth of the tree is $M=3$, the function $\varphi$ is defined by taking $\delta=1.1$. For the precision parameter we take a prior $\alpha\sim\ga(1,2)$ and for the location parameter of the centring measure we take $\mu_l\sim\no(0,\tau_{\mu})$ for $l=1,2,3$, with varying $\tau_{\mu}\in\{1,100\}$ to compare. Posterior inference was based on our adaptive MCMC with 10,000 iterations with a burn-in of 2,000 and keeping one of every 10$^{th}$ iteration after burn-in. Using an Intel Xenon at 2.20 GHz, 2 cores and 12 GB of RAM, it took around 45 minutes for each dataset. 

Table 1 shows the LPML and DIC measures obtained with the two prior specifications for the two datasets. For dataset B15 there is little difference between the fittings for the two different values of $\tau_{\mu}$, whereas for B19 there is a huge gain when taking $\tau_{\mu}=100$ as compared to $\tau_{\mu}=1$. So we take $\tau_{\mu}=1$, for B15, and $\tau_{\mu}=100$, for B19. We further report posterior inferences for some model parameters for the best fitting model. Posterior 95\% credible intervals (CI) for the precision parameter $\alpha$ are around $1$ for B15 and around $0.5$ for B19. Recall that the parameter vector $\bmu$ also plays the role of precision in the sense that if the magnitude $||\bmu||$ is large, then the projected P\'olya tree prior is highly concentrated around its mean. On the other hand, the prior on $\bmu$ induces a smoothness effect in the posterior density estimates. For the two datasets, posterior estimates of $\mu_2$ and $\mu_3$ are around zero (highest dispersion), but for $\mu_1$ posterior estimates are negative for B15 and positive for B19. To validate our results, we also fitted the $\PPT$ model with a depth of $M=4$, goodness of fit measures and inferences are almost identical. For comparison purposes, Table \ref{tab:real} also reports the gof values when fitting the parametric projected normal model. For both datasets our nonparametric model is superior. 

\begin{figure}
\centerline{\includegraphics[scale=0.45]{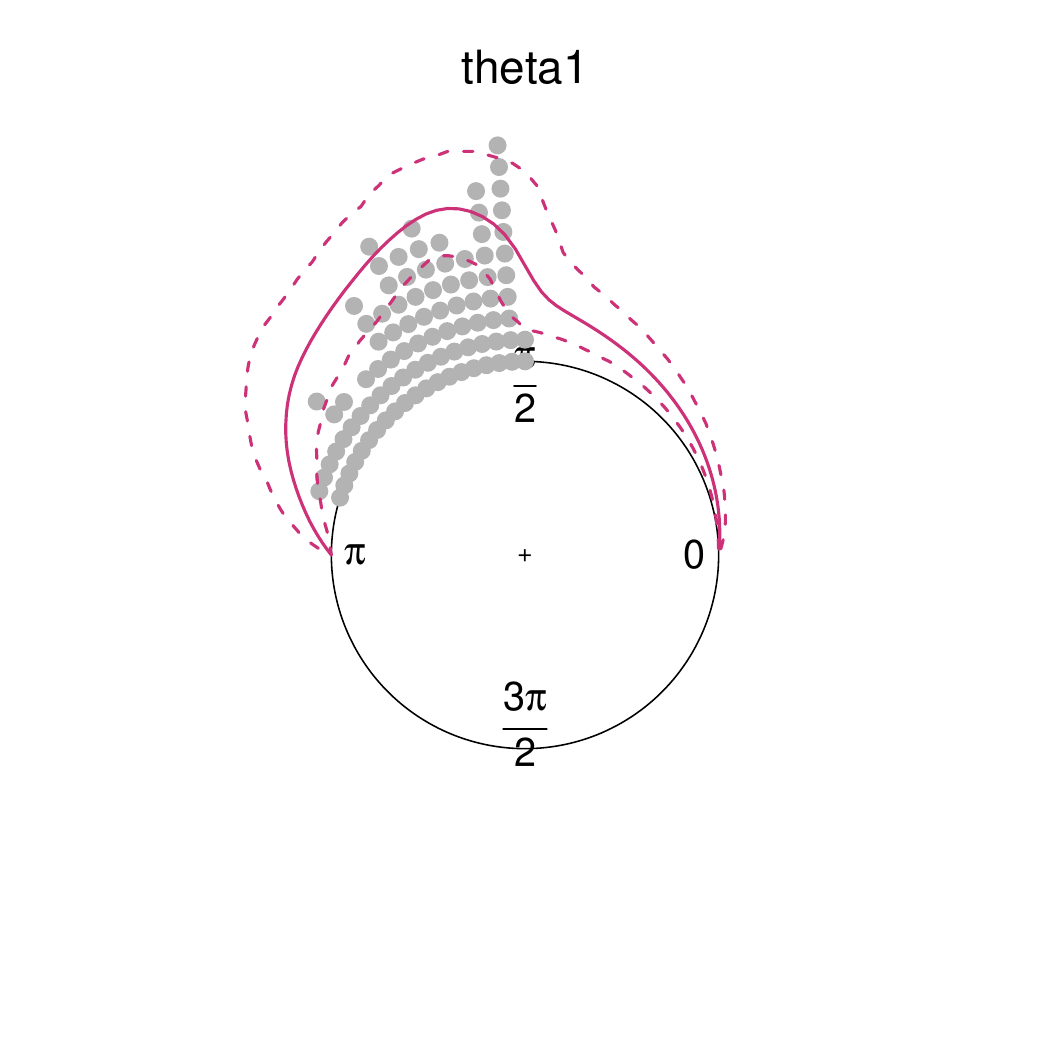}
\includegraphics[scale=0.45]{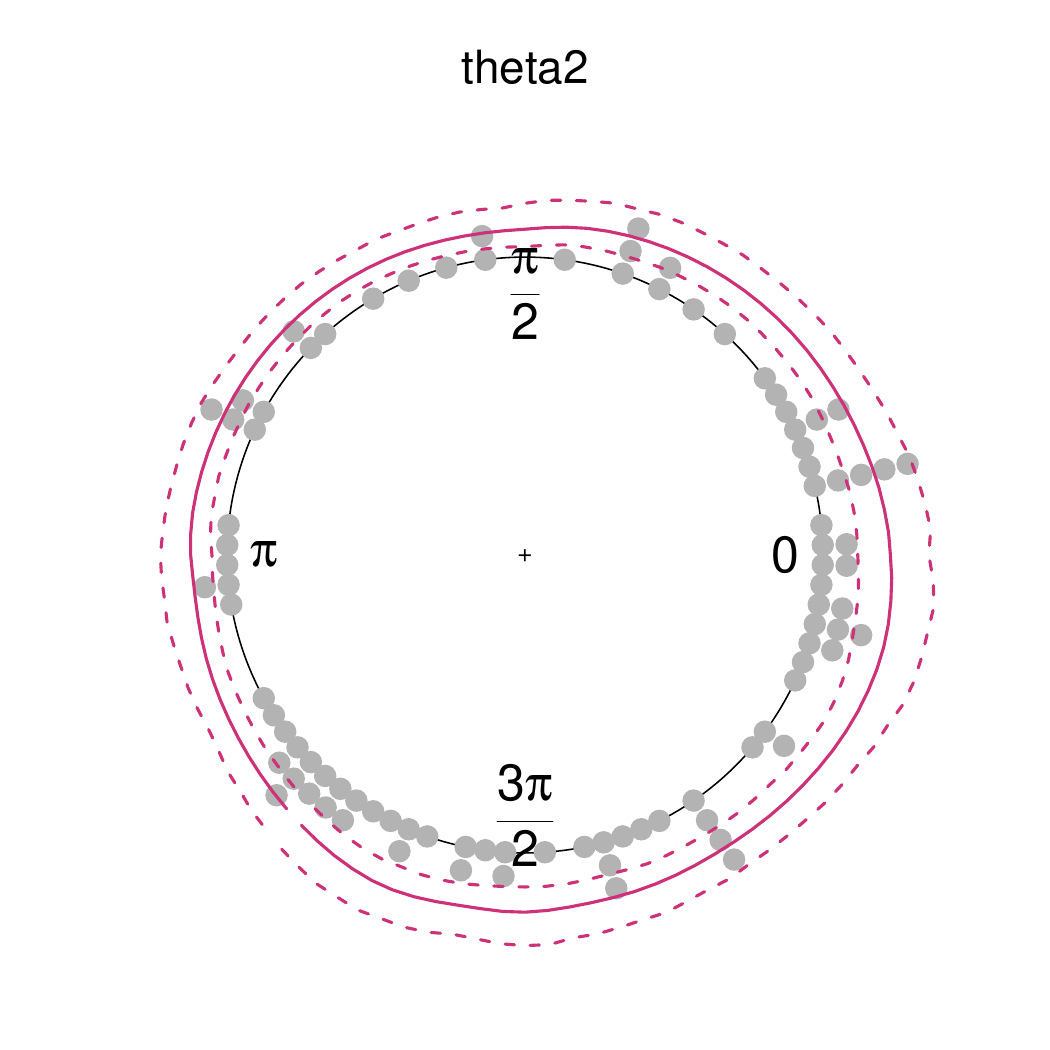}}
\centerline{\includegraphics[scale=0.45,angle=270,origin=c]{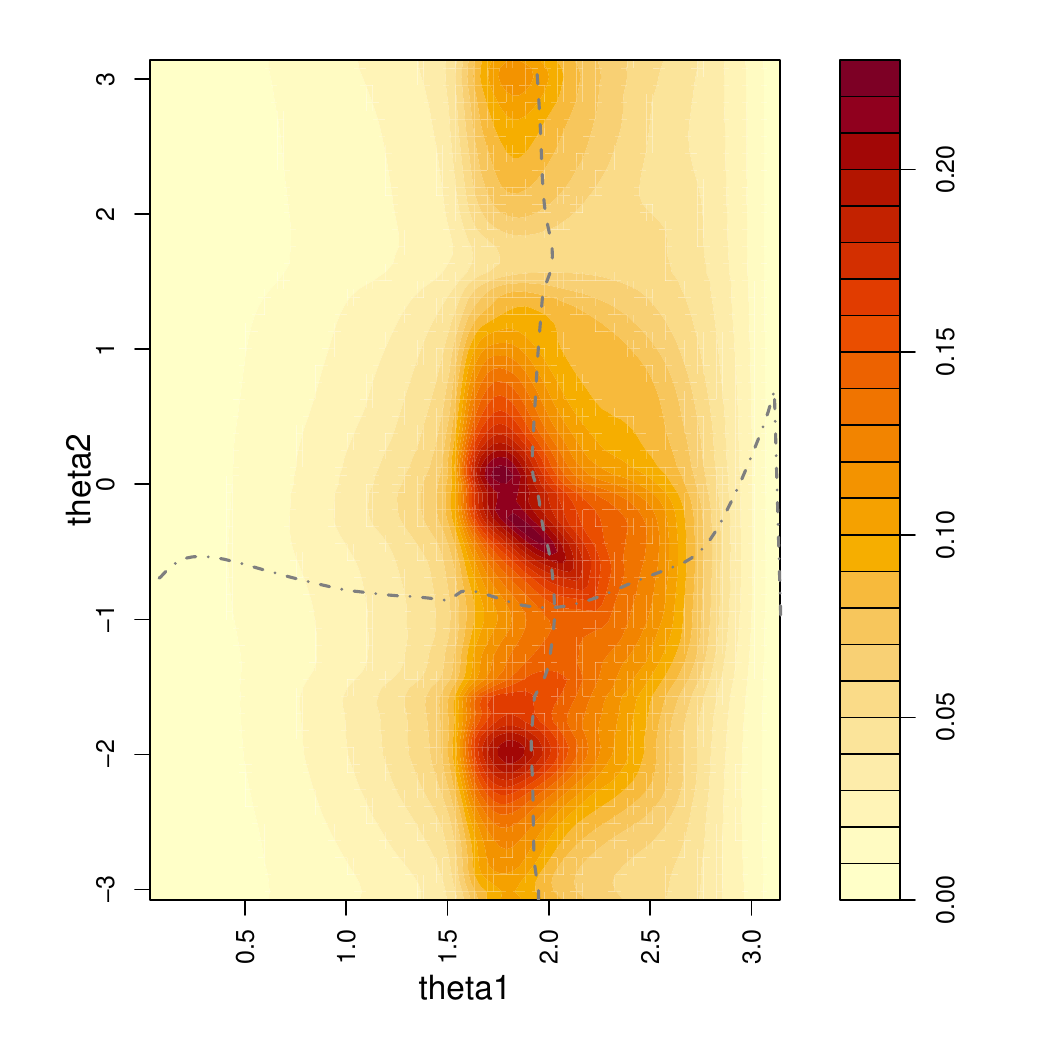}
\includegraphics[scale=0.45]{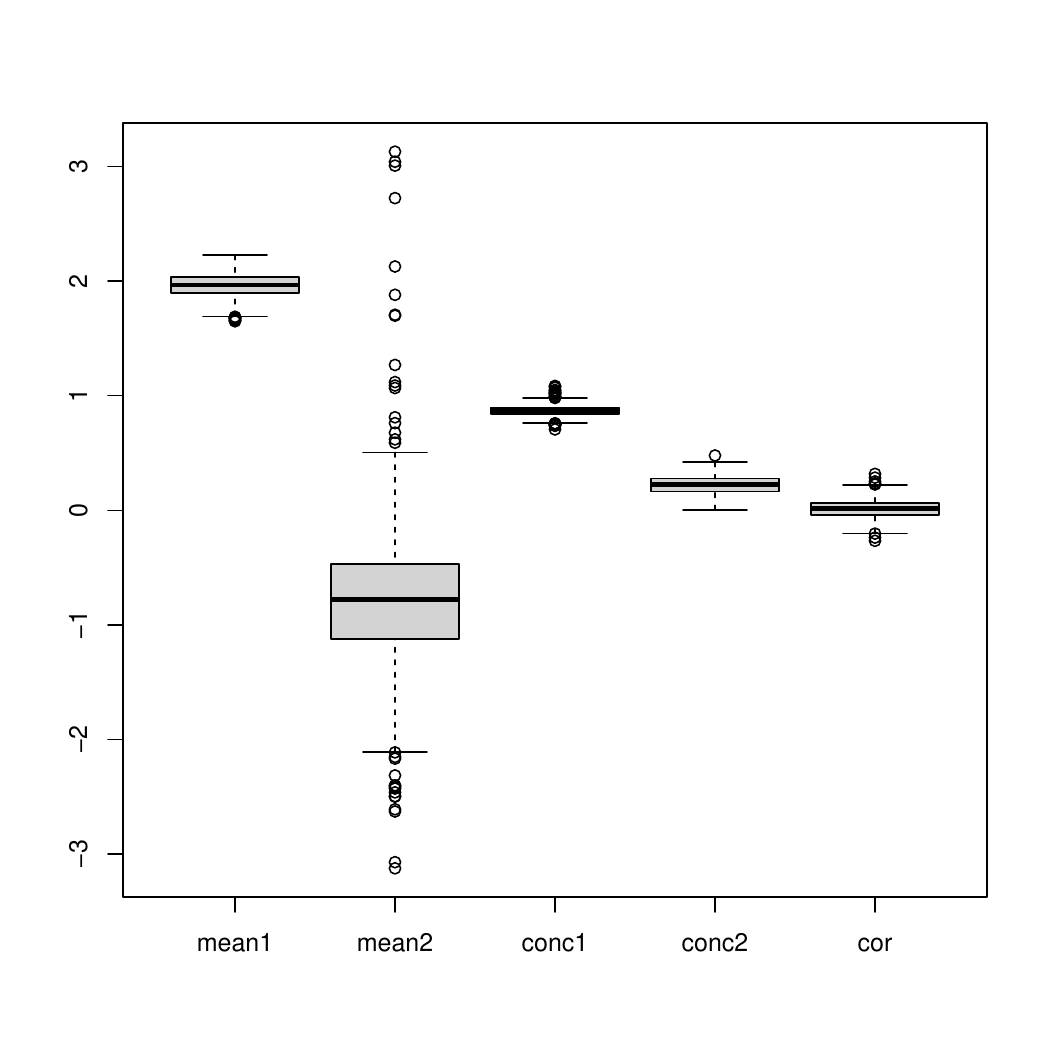}}
\caption{{\small Posterior inference for dataset B15. Marginal densities for $\Theta_1$ (top-left) and $\Theta_2$ (top-right), posterior mean (solid line) and 95\% CI (dotted line). Contour plot of joint density (bottom-left) and conditional means $\E(\Theta_1\mid\theta_2)$ (dashed line) and $\E(\Theta_2\mid\theta_1)$ (dashdotted line). Boxplot of moments (bottom-right).}}
\label{fig:realB15}
\end{figure}

\begin{figure}
\centerline{\includegraphics[scale=0.45]{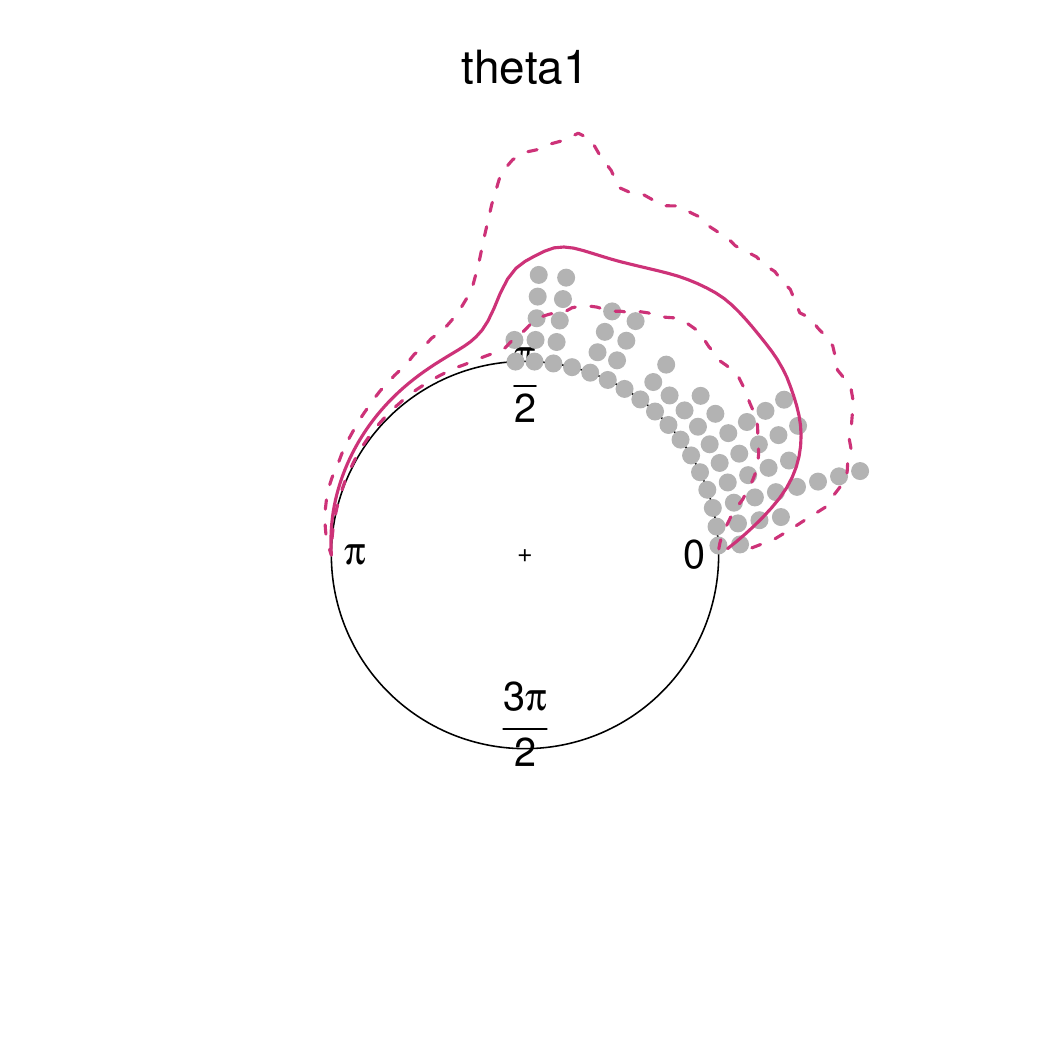}
\includegraphics[scale=0.45]{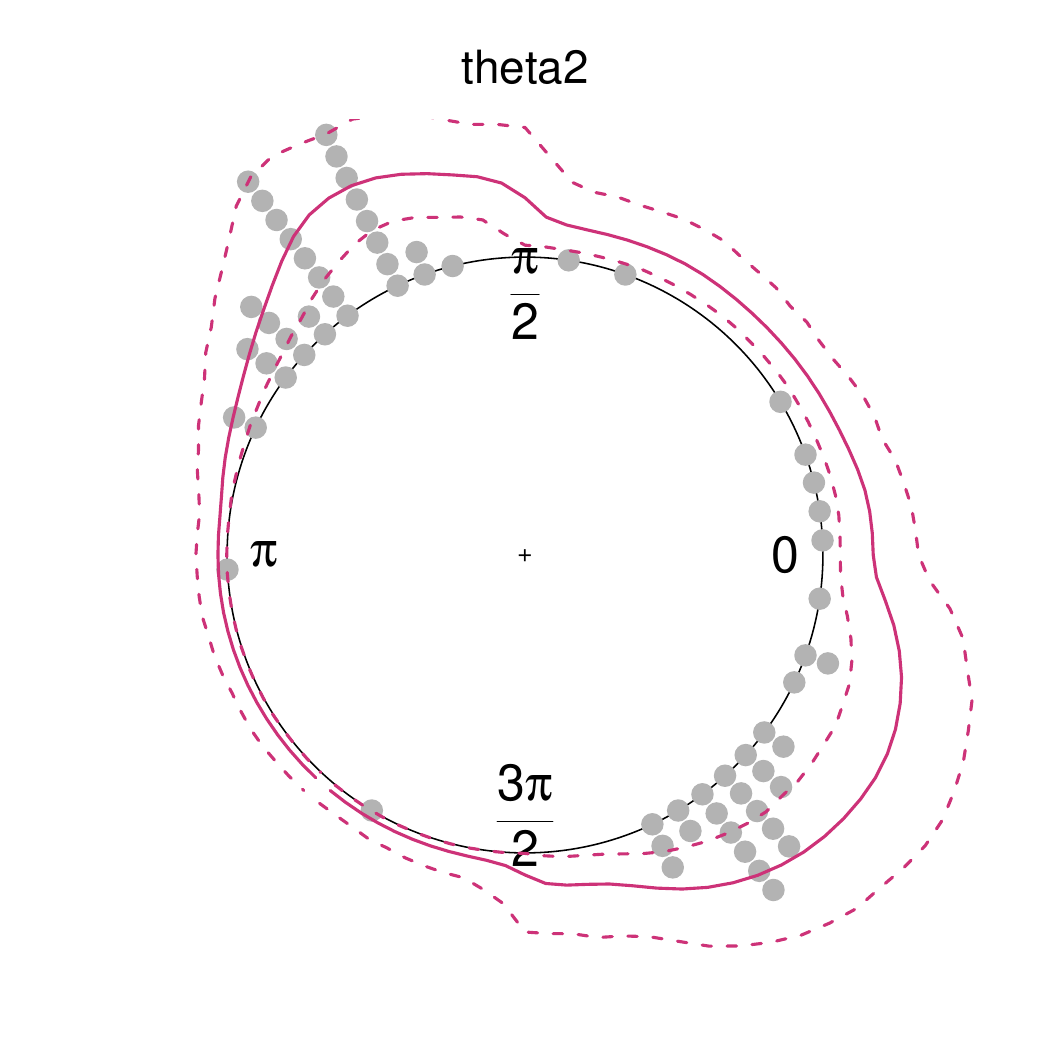}}
\centerline{\includegraphics[scale=0.45,angle=270,origin=c]{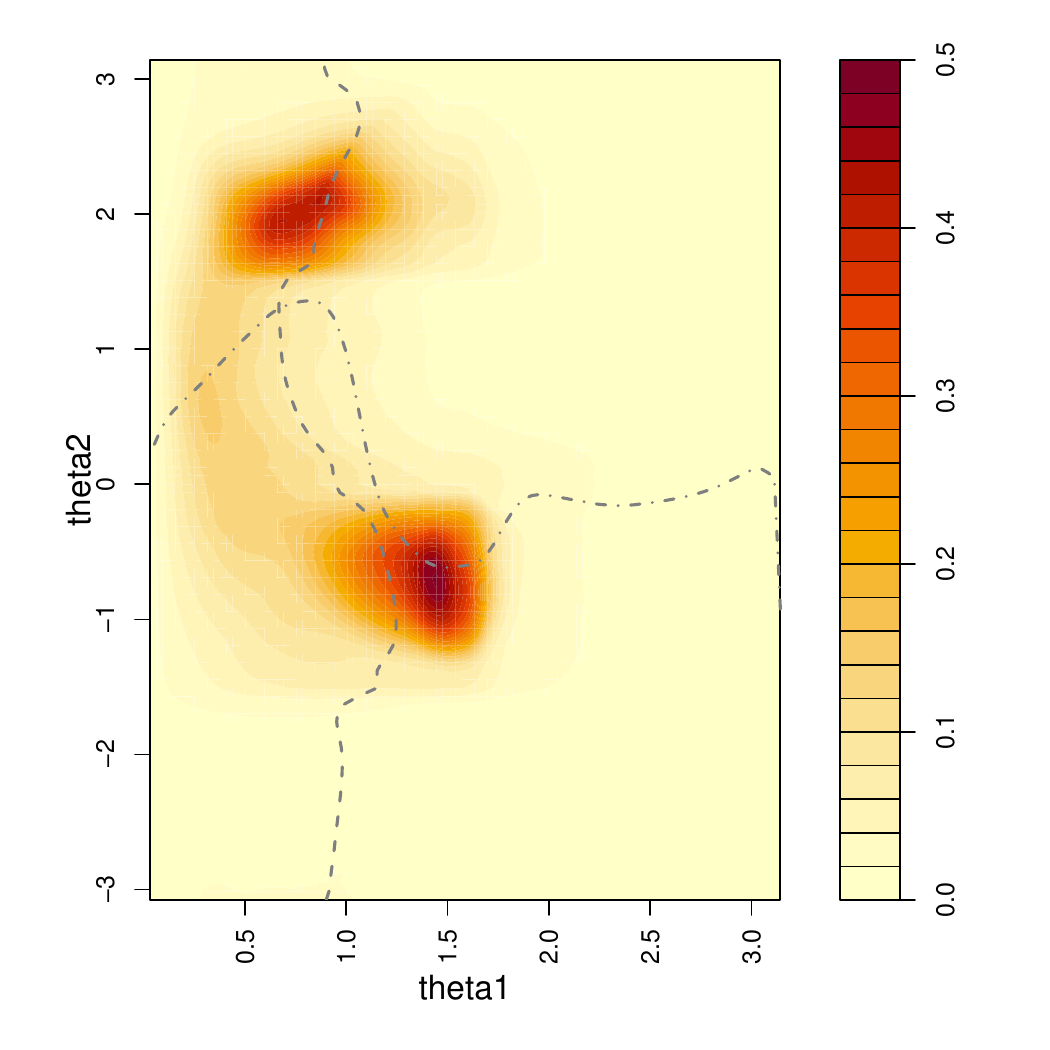}
\includegraphics[scale=0.45]{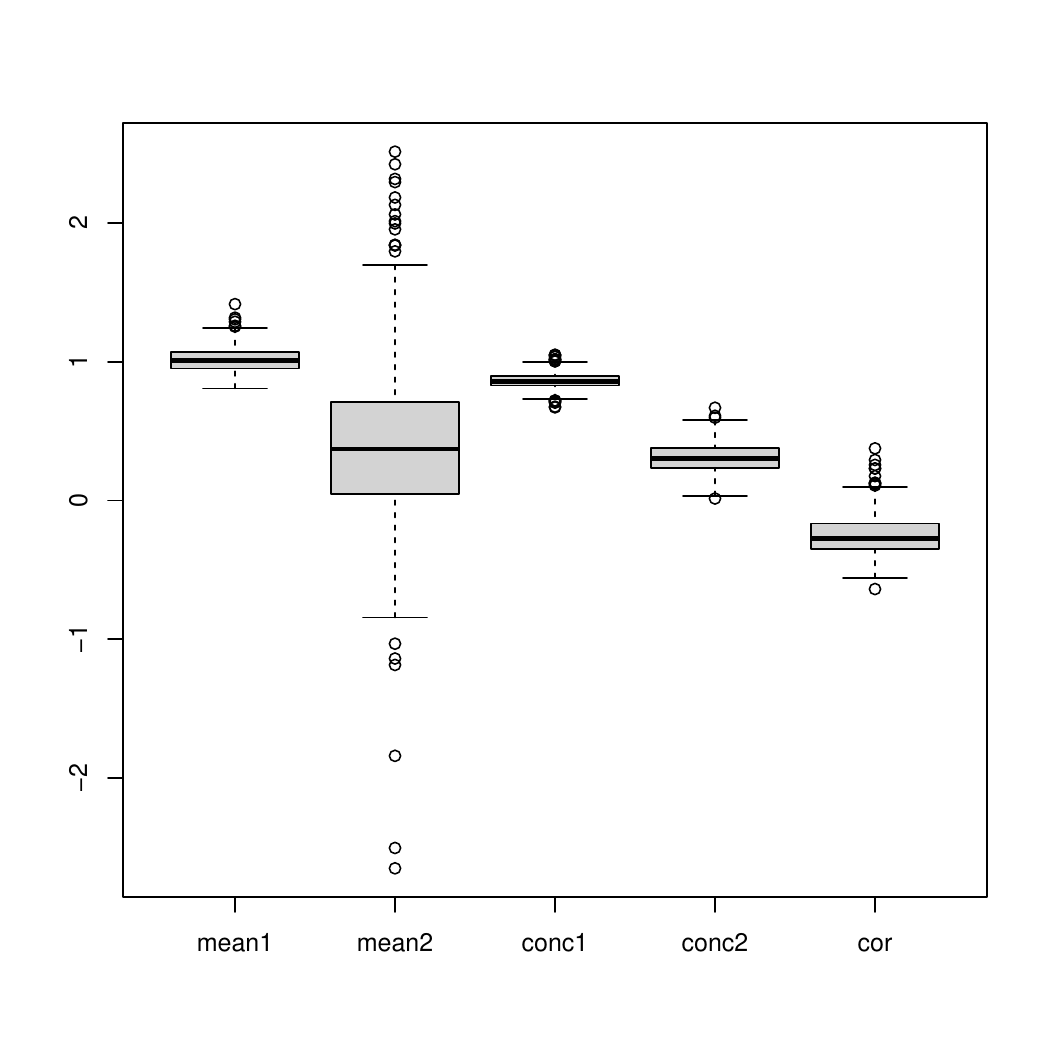}}
\caption{{\small Posterior inference for dataset B19. Marginal densities for $\Theta_1$ (top-left) and $\Theta_2$ (top-right), posterior mean (solid line) and 95\% CI (dotted line). Contour plot of joint density (bottom-left) and conditional means $\E(\Theta_1\mid\theta_2)$ (dashed line) and $\E(\Theta_2\mid\theta_1)$ (dashdotted line). Boxplot of moments (bottom-right).}}
\label{fig:realB19}
\end{figure}

Marginal, joint density estimates and moments are reported in Figures \ref{fig:realB15} and \ref{fig:realB19}, for each dataset, respectively. The observed values of the first coordinate $\Theta_1$ lie in the subinterval $[\pi/2,\pi]$ for B15, and in $[0,\pi/2]$ for B19. The second (circular) coordinate $\Theta_2$, in both datasets, takes values in the whole circle $[0,2\pi]$, where for B15 the distribution is more uniform, whereas for B19 the distribution is bimodal with opposite modes at $3\pi/4$ and $-3\pi/4$, respectively. For both datasets, marginal density estimates (top row) follow closely the shape of the data, but in a smooth way. 

For B15 (Figure \ref{fig:realB15}), marginal density estimate for $\Theta_1$ is unimodal with the mode around $2\pi/3$, whereas for $\Theta_2$ marginal density is somehow uniform, perhaps with two small modes around $-2\pi/3$ and $0$. Joint density estimate, shown as a heat map (bottom left panel), represents a 2-D projection of the eye of a cat and indicates two predominant modes (darker shadows) surrounding the lower part of the retina and perhaps one smaller mode around coordinates $(2\pi/3,\pi)$, which could be seen as the refraction of the light at the back of the eye. In any case, dendritic fields are just present in the southern hemisphere of the eye. Regression curves, $\E(\Theta_1\mid\theta_2)$ and $\E(\Theta_2\mid\theta_1)$, are also included as dashed and dotted lines, respectively, over the heat map. Since these two lines are almost parallel to the axes, we do not see any explanation power in any of the two explanatory variables. Regarding the moments, posterior distributions are included as box plots (bottom right panel). Posterior distributions of the mean directions have totally different behaviour, $\nu_1$ shows almost null dispersion, whereas $\nu_2$ has a lot more dispersion. Regarding concentrations, $\varrho_1>\varrho_2$. Linear dependence measure $\rho$ shows little dispersion and is highly concentrated around zero. In fact 95\% CI for $\rho$ is $(-0.14,0.19)$, confirming that there is almost no dependence between these two coordinates (latitude and longitude). 

For B19 (Figure \ref{fig:realB19}), observed data for $\Theta_1$ (grey dots) shows a slight higher concentration close to zero, however, posterior estimates of the marginal density for $\Theta_1$ is almost uniform. On the other hand, data and marginal density estimates for $\Theta_2$ both show a bimodality, with modes around $-\pi/3$ and $2\pi/3$. Joint density estimate shows two predominant modes around coordinates $(\pi/2,-\pi/3)$ and $(\pi/3,2\pi/3)$. Gross speaking, the heat map shows a horseshoe  shape for beddings, which suggests a non linear dependence between the two angles. This behaviour could not be uncovered by the projected normal model. Regression curves, $\E(\Theta_1\mid\theta_2)$ and $\E(\Theta_2\mid\theta_1)$, wiggle around the two modes. For instance, if $\theta_1<1.3$, mean values for $\Theta_2$ are positive (north-east direction), whereas if $1.3<\theta_1<1.8$, mean values for $\Theta_2$ are negative (south-east direction), and if $\theta_1>1.8$ mean values for $\Theta_2$ are around zero (east-direction). Complementing this, the box plot of dependence measure $\rho$ is mainly located in negative values, with posterior mean $-0.25$ and posterior probability of $\rho$ being negative $0.96$. This confirms a negative dependence between dip and dip direction in the folds. Moreover, marginal mean directions $\nu_1$ and $\nu_2$ are both positive around $\pi/3$ and $\pi/6$, respectively, and marginal concentrations behave similarly to dataset B15 with $\varrho_1>\varrho_2$.

\subsection{Real data analysis with covariates}

Considering again \cite{fisher&al:87}, dataset B23 consists of measurements of natural remanent magnetisation in Old Red Sandstones rocks from two sites, A and B, in Pembrokeshire Wales. The spherical observed data are inclination ($\eta_1$) and declination ($\eta_2$), with values in $[-25,52]$ and $[167,235]$, respectively. We transform the data into radians by defining $\theta_1=\pi(\eta_1+90)/180$ and $\theta_2=2\pi\eta_2/360$. We define indicator variables $Z_1=1$ if observation comes from site A and $Z_2=1$ if observation comes from site B. 

Available data therefore consists of $(\Theta_{1,i},\Theta_{2,i},Z_{1,i},Z_{2,i})$ for $i=1,\ldots,n$. Data size is $n=48$, where $35$ of them come from site A and $13$ from site B. For this dataset we fit our linear-directional regression model as the one described in Section \ref{sec:model}, with explanatory variables $\bZ$ of dimension $p=2$ and no intercept. The matrix of coefficients $\Gamma$ is of dimension $3\times 2$. Prior specifications for the P\'olya tree are as in the previous example,  $M=3$, $\delta=1.1$, $\alpha\sim\ga(1,2)$. For the regression coefficients we took independent prior distributions of the form $\gamma_{l,h}\sim\no(0,0.1)$. Posterior inference was based on our adaptive MCMC with 52,000 iterations with a burn-in of 20,000 and keeping one of every 40$^{th}$ iteration after burn-in. Computational time was around 90 minutes in a machine with the characteristics defined above. 

\begin{figure}
\centerline{\includegraphics[scale=0.45,angle=270,origin=c]{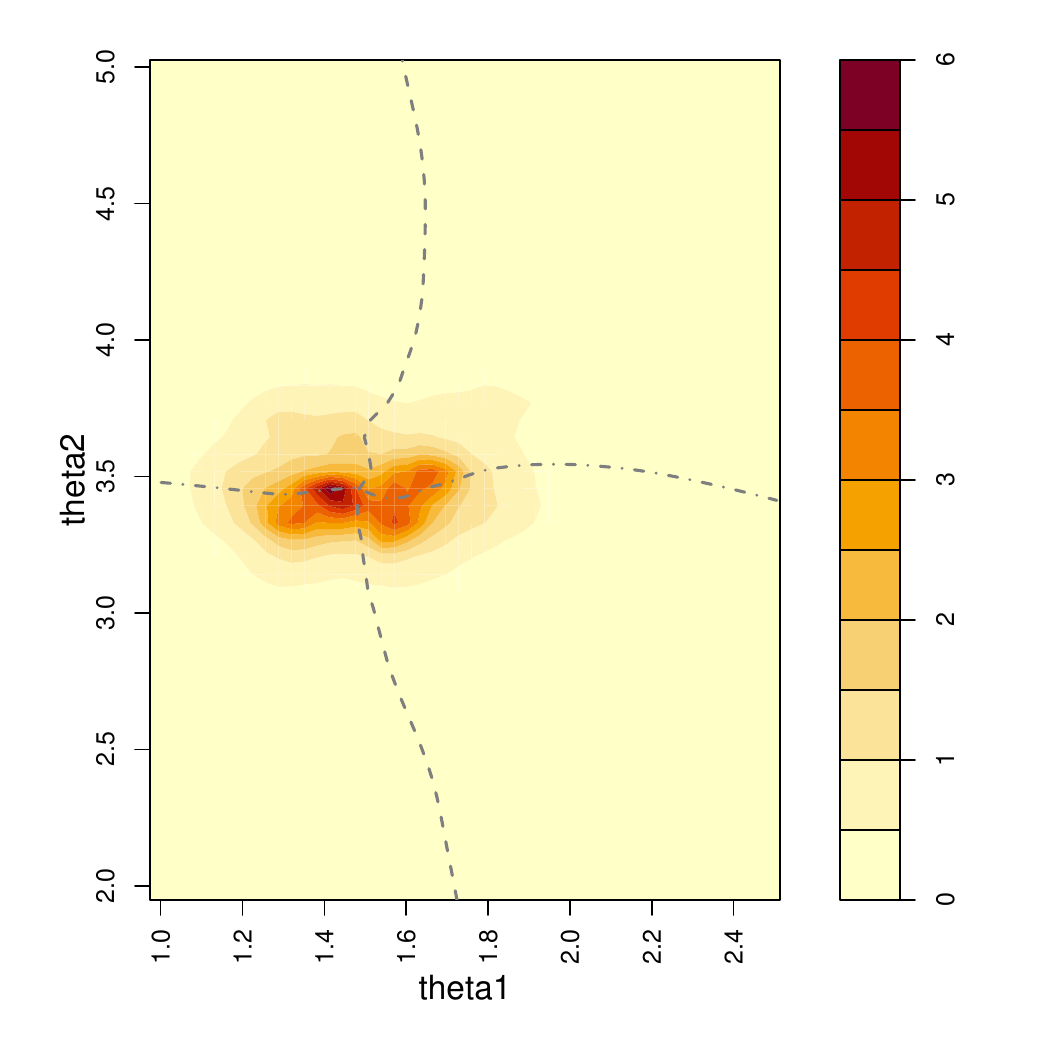}
\includegraphics[scale=0.45,angle=270,origin=c]{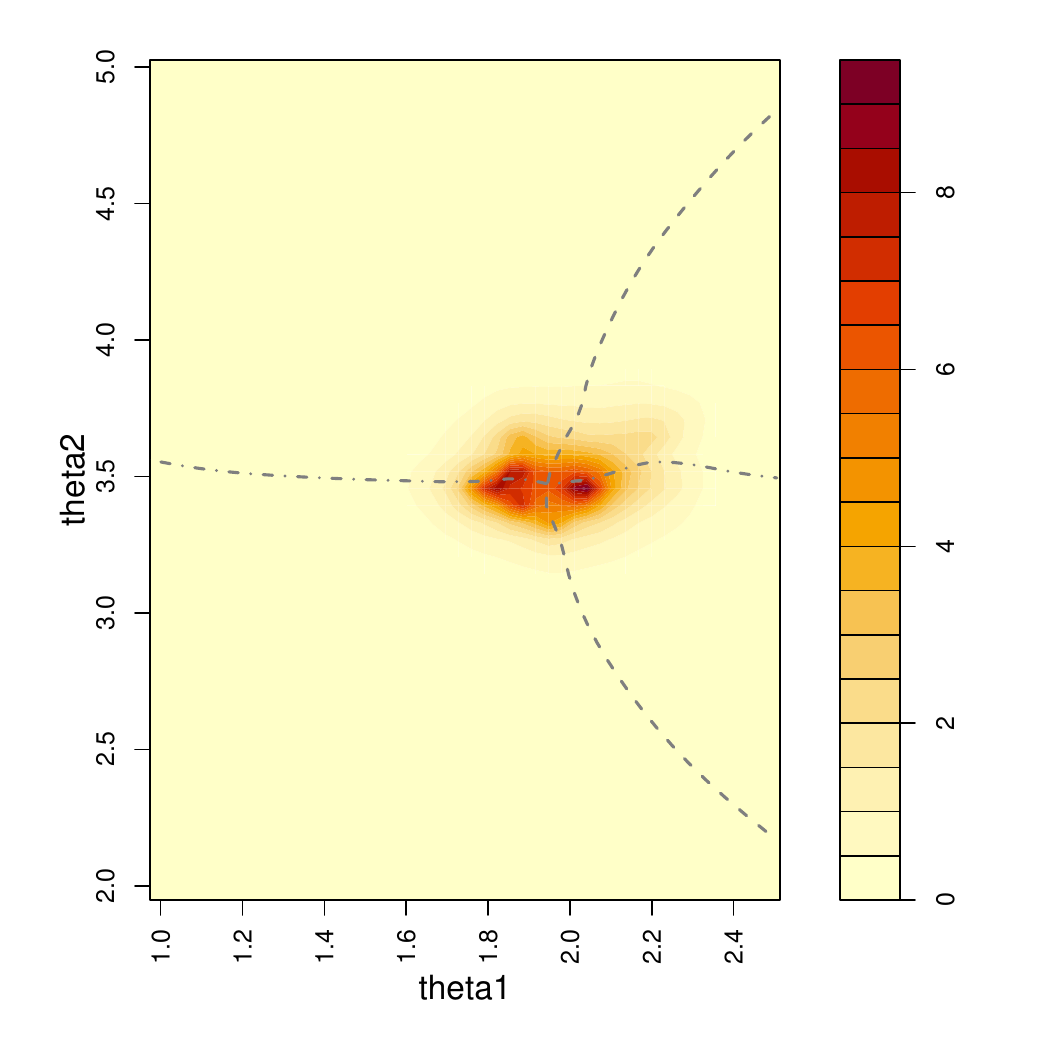}}
\caption{{\small Posterior inference for dataset B23. Contour plot of joint densities for site A (left) and site B (right). Conditional means $\E(\Theta_1\mid\theta_2)$ (dashed line) and $\E(\Theta_2\mid\theta_1)$ (dashdotted line).}}
\label{fig:realB23ab}
\end{figure}

Figure \ref{fig:realB23ab} displays the joint posterior predictive for sites A (left panel) and B (right panel). Note that in both graphs we did a close up around the region of high probability because the data are highly concentrated. The contours seem to be mainly parallel to the axis and the correlation between $\Theta_1$ and $\Theta_2$, in both sites, is almost null with point estimates $0.08$ and $0.11$, respectively. Moreover, $\P(\rho>0\mid\data)$ is $0.79$ and $0.85$ in sites A and B, respectively. This low association between angles is also shown by the conditional means (dashed and dotted lines), which are mainly parallel to the axis with slight curves around the areas of high probability. In particular, in site B, $\E(\Theta_1\mid\theta_2)$ (dashed line) shows a ``V'' shape in the center of the joint distribution. 

\begin{figure}
\centerline{\includegraphics[scale=0.7]{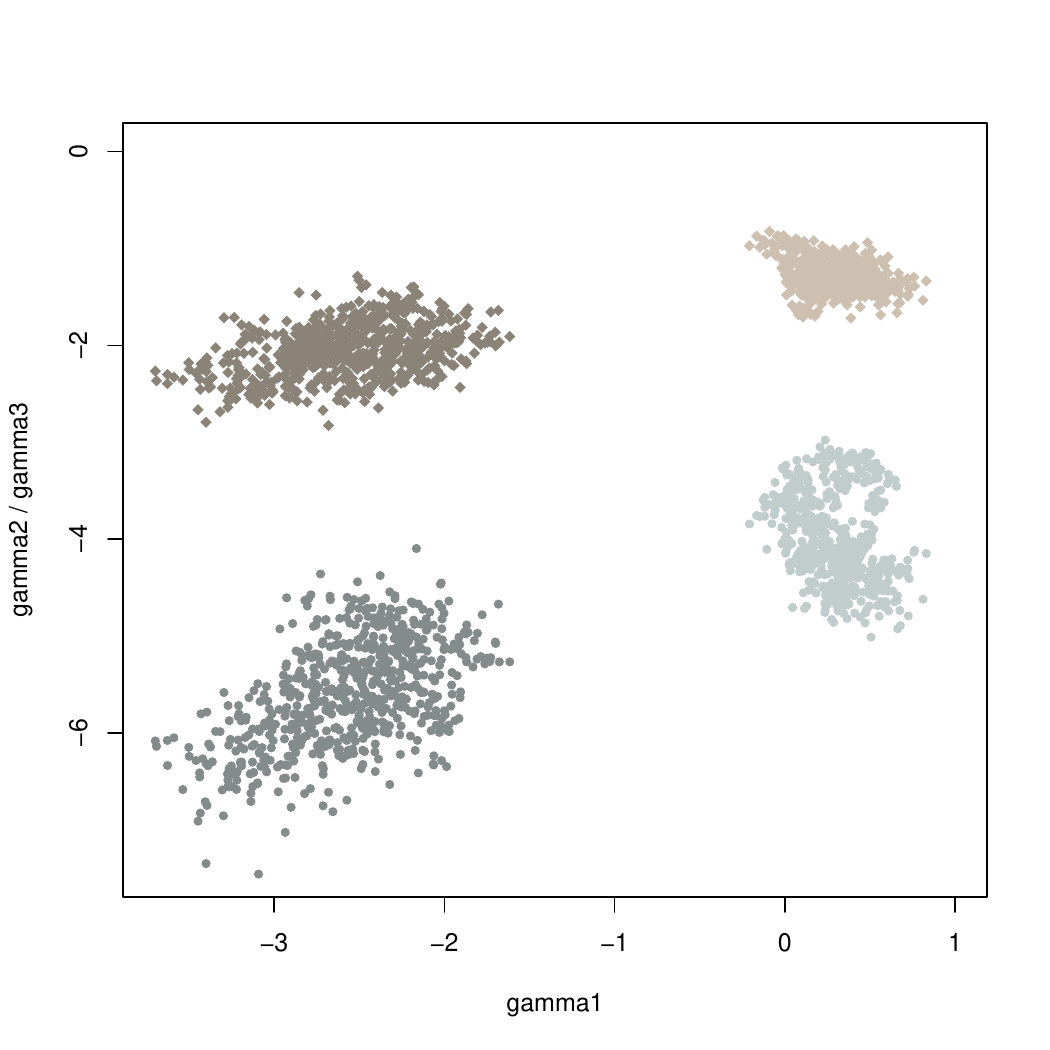}}
\caption{{\small Posterior inference for dataset B23. Dispersion diagrams of regression parameters: $\gamma_{1,1}$ vs $\gamma_{2,1}$ (light dots), $\gamma_{1,1}$ vs $\gamma_{3,1}$ (light squares), $\gamma_{1,2}$ vs $\gamma_{2,2}$ (dark dots), $\gamma_{1,2}$ vs $\gamma_{3,2}$ (dark squares).}}
\label{fig:realB23g}
\end{figure}

One apparent difference between the joint distributions of the remanent magnetisations in the two sites is a shift in the location of the inclination variable $\Theta_1$ which is located around 1.5 radians for site A and around 2 radians for site B. In order to understand how this shift is captured by our model, we show in Figure \ref{fig:realB23g} the joint posterior distributions of the regression coefficients $\gamma_{l,h}$, for $l=1,2,3$ and $h=1,2$, as dispersion diagrams. We first note that the coefficients are all away from zero with a marginal probability larger than $0.95$. Parameters associated to site A are shown as light colours in the figure and have posterior estimates $\hat\gamma_{1,1}=0.28$, $\hat\gamma_{2,1}=-3.76$ and $\hat\gamma_{3,1}=-1.21$. Parameters associated to site B are shown as dark colours in the figure and have posterior estimates $\hat\gamma_{1,2}=-2.49$, $\hat\gamma_{2,2}=-5.50$ and $\hat\gamma_{3,2}=-2.00$. Therefore, using first panel in Figure \ref{fig:coef}, we conclude that the shift in the location of $\Theta_1$ is controlled by $\gamma_{1,1}$ and $\gamma_{1,2}$. A positive value induces a smaller inclination (site A) and a negative value induces a larger inclination (site B).

\section{Concluding remarks}
\label{sec:concl}

We have proposed a Bayesian nonparametric model for directional data of any dimension. Our proposal is based on the projection of a multivariate P\'olya tree to the unit hypersphere. Random densities obtained from the model turned out to be smoother than typical densities arising from P\'olya trees, which are discontinuous at the boundaries of the partitions. Further smoothing can be obtained by taking hyper-prior distributions on the location parameter $\bmu$ of the centring distribution $F_0$. As a consequence of the joint multivariate modelling, we can define conditional distributions of some of the directions given the others and define directional-directional regression curves via conditional expected values. We emphasize that regression models defined in this way are not linear, since they are nonparametric, they can capture any linear and non linear relationship. Inclusion of linear covariates was also possible by considering median regression models for the unprojected directions. 

Posterior inference, for models with and without linear covariates, is done in a similar way. We just  augment the data with (unobserved) latent resultants and update the multivariate P\'olya tree. To simplify the posterior dependence on the prior choice of $\alpha$, we suggested to place a hyper-prior on this parameter, with minimal extra effort in sampling from its conditional posterior distribution. Adaptive MH steps were proposed to achieve optimal acceptance rates and improve performance. 

Although posterior inference is theoretically simple, it is computationally very demanding, specially for dimensions $k\geq 3$. Multivariate P\'olya trees are defined via nested partitions of $\RB^k$ into $2^{km}$ sets for each level $m=1,\ldots,M$. For $k=2$, values of $M\leq 4$ produce a manageable number of partitioning sets, $2^{kM}\leq 256$. For $k=3$, we can only go up to $M=3$ to produce posterior inferences in a reasonable amount of time because $2^{kM}\leq 512$. For larger values of $k$ and or $M$, inference becomes computationally intensive, due to the enormous amount of time needed. One possible solution is to define asymmetric partition depths, that is, for each coordinate $l$, define a different maximum depth, say $M_l$ for $l=1,\ldots,k$, such that the final partition size is $2^{\sum_{l=1}^k M_l}$. However it might not be easy to determine an optimal maximum depth for each coordinate because we do not actually observe data in $\RB^k$, we observe data in the hypersphere $\SB^k$, which is characterised by $k-1$ directions and augment them with an extra latent resultant to produce data in the whole space $\RB^k$, so we can not easily identify coordinates that would require finer partitions. 

However, the most time consuming task is the numerical evaluation of the density $f(\btheta)$ because for every dimension $l=1,\ldots,k-1$ we use an evaluation grid of $T$ points giving a total of $T^{k-1}$ points. For each of these points we do a quadrature numerical integration with another grid of size $T$. Therefore the computational burden is of order $T^k$. For the examples considered here we took $T=100$ and $k=3$ giving a total of one million evaluations every time we want to compute the joint density. 

Proposing solutions to solve the computational burden, due to the curse of dimensionality, is open and left for further study in a future work.

\section*{Acknowledgements}

The author acknowledges support from \textit{Asociaci\'on Mexicana de Cultura, A.C.} and is grateful for comments from anonymous referees.

\newpage

\begin{table}
\caption{Simulation study. Goodness of fit measures for projected normal and PPT.}
\label{tab:sim}
\begin{center}
\begin{tabular}{cccccc} \hline \hline
 & & \multicolumn{2}{c}{S1} & \multicolumn{2}{c}{S2} \\ \hline
Model & $n$ & LPML & DIC & LPML & DIC \\ \hline
Proj.Normal & $100$ & $-211.0$ & $421.9$ & $-296.7$ & $592.9$ \\ 
$\PPT_2$ & $100$ & $-230.7$ & $459.2$ & $-230.9$ & $371.1$ \\
Proj.Normal & $300$ & $-666.7$ & $1333.4$ & $-892.8$ & $1784.9$ \\ 
$\PPT_2$ & $300$ & $-691.5$ & $1380.3$ & $-524.4$ & $999.7$ \\
\hline \hline
\end{tabular}
\end{center}
\end{table}

\begin{table}
\caption{Real data analysis. Goodness of fit measures for projected normal and PPT with different values of $\tau_{\mu}$, together with 95\% credible intervals (CI) for some model parameters for best model.}
\label{tab:real}
\begin{center}
\begin{tabular}{ccccc} \hline \hline
 & \multicolumn{2}{c}{B15} & \multicolumn{2}{c}{B19} \\ \hline
Model & LPML & DIC & LPML & DIC \\ \hline
Proj.Normal & $-238.6$ & $477.4$ & $-182.8$ & $365.5$ \\ 
$\PPT_2(\tau_{\mu}=1)$ & $-232.2$ & $454.0$ & $-217.1$ & $391.7$ \\
$\PPT_2(\tau_{\mu}=100)$ & $-233.6$ & $473.7$ & $-199.9$ & $261.3$ \\ \hline
Parameters & \multicolumn{4}{c}{95\% CI} \\ \hline
$\alpha$ & \multicolumn{2}{c}{$(0.57,2.50)$} & \multicolumn{2}{c}{$(0.16,0.64)$} \\
$\mu_1$ & \multicolumn{2}{c}{$(-0.92,-0.48)$} & \multicolumn{2}{c}{$(0.26,0.64)$} \\
$\mu_2$ & \multicolumn{2}{c}{$(-0.08,0.36)$} & \multicolumn{2}{c}{$(-0.15,0.16)$} \\
$\mu_3$ & \multicolumn{2}{c}{$(-0.47,0.01)$} & \multicolumn{2}{c}{$(-0.13,0.19)$} \\
\hline \hline
\end{tabular}
\end{center}
\end{table}

\end{document}